\documentclass[aps,a4paper,superscriptaddress,twocolumn]{revtex4}
\usepackage{tensor}
\usepackage{graphicx}
\usepackage{amsmath}
\usepackage{amssymb}
\usepackage{enumerate}
\usepackage{subfigure}
\usepackage{tabularx}
\usepackage[colorlinks=true, pdfstartview=FitV, linkcolor=blue, citecolor=red, urlcolor=black, breaklinks=true]{hyperref}
\newcommand{\be}{\begin{equation}}
\newcommand{\ee}{\end{equation}}
\newcommand{\ben}{\begin{eqnarray}}
\newcommand{\een}{\end{eqnarray}}
\newcommand{\bes}{\begin{subequations}}
\newcommand{\ees}{\end{subequations}}
\def\bal#1\eal{\begin{align}#1\end{align}}

\newcommand{\sech}{{\rm sech}}
\newcommand{\LL}{{\mathcal L}}

\begin{document}
\title{Electrically charged localized structures}
\author{D. Bazeia}
\affiliation{Departamento de F\'\i sica, Universidade Federal da Para\'\i ba, 58051-970 Jo\~ao Pessoa, PB, Brazil}
\author{M.A. Marques}
\affiliation{Departamento de Biotecnologia, Universidade Federal da Para\'\i ba, 58051-900 Jo\~ao Pessoa, PB, Brazil}
\author{R. Menezes}
\affiliation{Departamento de Ci\^encias Exatas, Universidade Federal
da Para\'{\i}ba, 58297-000 Rio Tinto, PB, Brazil}
\affiliation{Departamento de F\'\i sica, Universidade Federal da Para\'\i ba, 58051-970 Jo\~ao Pessoa, PB, Brazil}
\begin{abstract}
This work deals with an Abelian gauge field in the presence of an electric charge immersed in a medium controlled by neutral scalar fields, which interact with the gauge field through a generalized dielectric function. We develop an interesting procedure to solve the equations of motion, which is based on the minimization of the energy, leading us to a first order framework where minimum energy solutions of first order differential equations solve the equations of motion. We investigate two distinct models in two and three spatial dimensions and illustrate the general results with some examples of current interest, implementing a simple way to solve the problem with analytical solutions that engender internal structure.   
\end{abstract}
\date{\today}
\maketitle

\section{Introduction}

Localized finite energy structures play an important role in nonlinear science in general. In high energy physics, in particular, localized structures may attain topological profile and may appear as kinks in the real line, vortices in the plane and monopoles in the three dimensional space \cite{B1,B2}. Kinks can be immersed in the plane as domain ribbons and in space as domain walls, and vortices can behave as stringlike objects when immersed in the space. These structures are well-known objects and have been studied in several distinct contexts in high energy physics, and in applications in several areas of nonlinear science. 

In this work we will study localized structures, but following another route. The study will deal with systems composed of a single charge immersed in a medium with electric permittivity controlled by real scalar fields that regularize the energy of the system. Since the scalar field is electrically neutral, the coupling with the Abelian gauge field is via the Maxwell term, as it appears in the Lagrange density \eqref{lmodel}, for instance. This is not a standard coupling, but it has been used in several distinct situations, in particular in the so called Friedberg-Lee model \cite{FL1,FL2}, where a dielectric function is used to describe a bag similar to the MIT \cite{mit} and SLAC \cite{slac} bag models. This coupling has also appeared in \cite{vor1,vor2,vor3}, in connection with the presence of vortices in the plane. It was also explored in connection with the AdS/CFT correspondence, to describe insulators and metals within the holographic setup \cite{H1,H2}. The holographic scenario has been further explored in \cite{H3} and, in the context of the gauge/gravity duality, it was also used to investigate hydrodynamic behavior in a hot and dense strongly coupled relativistic fluid in the framework of the Einstein-Maxwell-dilaton theory \cite{H4}.

The use of a field-dependent function coupled with the gauge field dynamical term has also been recently considered in \cite{global} in the electric context, that is, in the presence of an electric charge fixed at the origin; there, we showed that the electric field has a behavior that captures the basic feature of asymptotic freedom, an effect that is usually associated to quarks and gluons. It was also considered, for instance, in \cite{internal,melni,research,nonminimal,Casa} to describe vortex configurations with internal structures in the plane in the magnetic context. Moreover, very recently, in \cite{adam} the authors studied the dielectric Skyrme model, that is, the Skyrme model where both the kinetic and Skyrme terms are multiplied by field-dependent functions, leading to new results of current interest. As one knows, the Skyrme model has a direct connection with mesons and baryons \cite{S1} and the binding energies of nuclei \cite{S2} and, in the same line, in \cite{gud} another investigation has been implemented, with focus on lowering the binding energies of the Skyrme solutions towards more realistic values. See also \cite{A} for the investigation of exact self-dual skyrmions, and \cite{B} for the construction of Lorentz invariant compactlike structures.  

  The field-dependent function can be used to modify the magnetic properties of the medium to study vortices, as in \cite{vor1,vor2,vor3,internal,melni,research,nonminimal,Casa}, but it can also be considered to modify electrical properties of the medium, as in \cite{FL1,FL2} and in \cite{global}, for instance. Inspired by the results of \cite{global}, in the present work we first consider the field-dependent function as a dielectric function to be governed by a single scalar field. In the sequence, we add an extra scalar field, that modifies both the dielectric function and the dynamical term of the first scalar field in the Lagrange density. We develop a first order framework that helps us to describe the system via first order differential equations, and show that it is valid for the two models. Also, in the case of two scalar fields the first order framework helps us to describe novel configurations that support electrically charged multilayered ringlike structures. An important feature of the first order framework is that it allows the presence of minimum energy field configurations that are solved analytically. The present study will focus on the basic aspects of the problem, which concerns the construction of electrically charged localized structures described by scalars and the electric field. However, the electrically charged structures which we report in the present work may also be of interest to other areas of physics, in particular, to optical fibers \cite{F} when the system supports axial symmetry, and to the study of memory and other devices based on ferroelectricity \cite{advances,science}.
    
  To implement the investigation, we organize the work as follows. In Sec. \ref{M} we first describe and illustrate the case with a single scalar field, and then introduce another model, described by two scalar fields, with is also illustrated with some distinct examples. Since we will be dealing with electrically charged localized structures, and since the electric field is a vector, the passage from two to three space dimensions is somehow smooth and requires no extra degrees of freedom, so we will study the case of two and three spatial dimensions in the present work, which are of potential  applications with dielectric and ferroelectric materials. The electric case is different from the case of magnetic structures, which may describe vortices in the plane or magnetic monopoles in space; since the magnetic field is pseudo vector, the passage from two to three spatial dimensions requires the inclusion of extra degrees of freedom, suggesting that we change from the Abelian $U(1)$ symmetry in the plane in the case of vortices \cite{nielsen}, to the non Abelian $SU(2)$ symmetry, for instance, when one moves on to the case of monopoles in three spatial dimensions \cite{M1,M2}. We then close the work in Sec. \ref{C}, reviewing the main results and adding comments on some new lines of investigation of current interest. 
  
 \section{The models}
 \label{M}

The investigation will focus on the presence of electrically charged localized structures in two and three spatial dimensions, so we split this Section in two distinct parts, the first dealing with the case of two spatial dimensions, and the second one with the case of three space dimensions. We will study two models, which are described by two distinct Lagrange densities that do not change when we go from two to three spatial dimensions. For this reason, we start describing the two models on general grounds, before specializing in the cases of two and three spatial dimensions.

We first consider the model with Lagrange density
\be\label{lmodel}
	\LL _1= - \frac{\varepsilon(\phi)}{4}F_{\mu\nu}F^{\mu\nu} +\frac12\partial_\mu\phi\partial^\mu\phi - A_\mu j^\mu,
\ee
where $A_\mu$ describes the Abelian gauge field, $F_{\mu\nu}=\partial_\mu A_\nu-\partial_\nu A_\mu$ stands for the electromagnetic strength tensor, $\phi$ is a real scalar field and $j^\mu$ is an external current density. $\varepsilon (\phi)$ is a real nonnegative function which only depends on $\phi$. It represents a dielectric function that couples the neutral real scalar field to the Abelian gauge field. Here we consider natural units, with $\hbar=c=1$ and more, we take time, space, fields and coupling constants dimensionless, for simplicity. Also, we will consider a unity electric charge, that is, $e=1$.

If the scalar field is constant and uniform, the model reduces to the case of an Abelian gauge field generated by the external current $j^\mu$. To circumvent this possibility, we use the scalar field to get to models of current interest. Moreover, we consider the case where the current density is a timelike vector of the form $j^\mu=(j^0,\vec{j}=0)$ with $j^0$ time-independent. In the present investigation we will then focus on how the scalar field may contribute to introduce modifications in the standard scenario.      

The equations of motion associated to the Lagrange density \eqref{lmodel} are
\bes\label{eomgen}
\bal
\partial_\mu \partial^\mu \phi + \frac{1}{4}\varepsilon_{\phi}F_{\mu\nu}F^{\mu\nu} &= 0, \\
\partial_\mu\left(\varepsilon F^{\mu\nu}\right) &= j^\nu.
\eal
\ees
where $\varepsilon_{\phi} = \partial \varepsilon/\partial\phi$. We  will search for static field configurations, and since there are no spatial components of the current density, there is no magnetic field present in the system. We then define the components of electric field $\textbf{E}$ as $E^i = F^{i0}$ to show that the above equations become
\bes
\bal\label{eomphi}
\nabla^2\phi + \frac12 \varepsilon_\phi|\textbf{E}|^2 &= 0, \\ \label{meqs}
\nabla \cdot (\varepsilon\,\textbf{E}) &= j^0.
\eal
\ees
The last equation is the Gauss' law of the model. The energy density associated to the solutions of the above equations is calculated standardly; it has the form
\be\label{rho}
\rho = \frac12\left(\nabla\phi\right)^2 + \frac12\varepsilon (\phi)|{\bf E}|^2 + A_0j^0.
\ee
The energy is found by integrating the above energy density.

The second model enlarges the above model, introducing an additional scalar field, $\chi$, to help control the dynamics of the gauge and the scalar field $\phi$. It is defined as
\be\label{lmodel2}
	\LL _2 = - \frac{\varepsilon(\phi,\chi)}{4}F_{\mu\nu}F^{\mu\nu}+ \frac{1}{2}f(\chi)\partial_\mu\phi\partial^\mu\phi + \partial_\mu\chi\partial^\mu\chi - A_\mu j^\mu,
\ee
where $\varepsilon(\phi,\chi)$ and $f(\chi)$ are real non negative functions. Notice that the dielectric function now depends on both $\phi$ and $\chi$ in this new model. The presence of the second scalar field $\chi$ is inspired on the recent work \cite{Liao}, in which the function $f(\chi)$ is introduced to modify the kinematics of the $\phi$ field. As we will see, this new model will bring novelties, modifying the internal struture of the field configurations.

In this new model, the equations of motion are
\bes
\bal
\partial_\mu \left(f\partial^\mu \phi\right) + \frac{1}{4}\varepsilon_{\phi}F_{\mu\nu}F^{\mu\nu} &= 0, \\
\partial_\mu\partial^\mu \chi- \frac12f_\chi\partial_\mu\phi\partial^\mu\phi+\frac{1}{4}\varepsilon_{\chi}F_{\mu\nu}F^{\mu\nu} &= 0,\\
\partial_\mu\left(\varepsilon F^{\mu\nu}\right) &= j^\nu.
\eal
\ees
In the case of static solutions we get
\bes
\bal
\nabla\cdot\left(f\nabla \phi\right) + \frac{1}{2}\varepsilon_{\phi}|{\bf E}|^2 &= 0, \\
\nabla^2\chi- \frac12f_\chi\left(\nabla\phi\right)^2+\frac{1}{2}\varepsilon_{\chi}|{\bf E}|^2 &= 0,\\
\nabla \cdot (\varepsilon\,\textbf{E}) &= j^0.
\eal
\ees
We can also write the energy density in the form
\be
\rho = \frac12f(\chi)\left(\nabla\phi\right)^2+ \frac12\left(\nabla\chi\right)^2 + \frac12\varepsilon (\phi,\chi)|{\bf E}|^2 + A_0j^0.
\ee
The above results are general and can be used in two and three spatial dimensions, with the coordinate systems which we consider below.
\subsection{Two Spatial dimensions}
\label{sec:plane}

 In the plane, we then consider polar coordinates $(r,\theta)$, for convenience, since we are interested in studying a single charge immersed in a medium with modified dielectric function.  In two spatial dimensions, the charge is represented by $j^0=\delta(r)/r$.
 
  \subsubsection{First model}
 
From Eq.~\eqref{meqs}, we obtain the electric field
\be\label{esingle}
\textbf{E} = \frac{1}{\varepsilon(\phi)} \frac{\hat{r}}{r}.
\ee
Also, knowing that the scalar electric fields do not depend on $\theta$, Eq.~\eqref{eomphi} can be rewritten as
\be
\frac1r\frac{d}{d r}\left(r\frac{d\phi}{d r}\right) +\frac12 \varepsilon_\phi|\textbf{E}|^2=0.
\ee
In this case, both $\textbf{E}$ and $\phi$ do not depend on $\theta$, and so the above equation becomes, after using the expression for the electric field in Eq.~\eqref{esingle},
\be\label{eom1}
r\frac{d}{dr}\left(r\frac{d\phi}{dr}\right) = \frac{d}{d\phi}\left(\frac{1}{2\varepsilon}\right).
\ee
This is a second order differential equation with nonlinearities introduced by the dielectric function $\varepsilon(\phi)$. 

We now follow the lines of Ref.~\cite{bogo} to find a first-order formalism for the above model. One can use Eq.~\eqref{rho} to write the energy density as $\rho = \rho_f + \rho_c$, where
\bes
\bal\label{rhofpolar}
	\rho_f &=\frac12\left(\frac{d\phi}{dr}\right)^2 + \frac{1}{2r^2\varepsilon(\phi)},\\
	\rho_c	 &= A_0\frac{\delta(r)}{r}.
\eal
\ees
with the indices $f$ and $c$ standing for field and charge, respectively. The total energy can be written in the form $E=E_f+E_c$, as the integral of the above energy densities, respectively. 

The key point here is that one can introduce an auxiliary function $W=W(\phi)$ to write $\rho_f$ in the form
\be
\rho_f = \frac12\left( \frac{d\phi}{dr} \mp \frac{W_\phi}{r}\right)^2 + \frac{1}{2r^2\varepsilon(\phi)} -\frac{W^2_\phi}{2r^2} \pm \frac1r\frac{dW}{dr}.
\ee
Notice the presence of the factor $1/r$ in the last term; it is important since under integration to get the energy, the line element  $dx\,dy=r\,dr\,d\theta$ makes $dW/dr$ be a surface term after integration. When the dielectric function is given by
\be\label{psingle}
\varepsilon(\phi) = \frac{1}{W_\phi^2},
\ee
one can show that the energy $E_f$ is bounded, i.e., $E_f\geq E_B$, where
\be\label{ebsingle}
E_B = 2\pi |W(\phi(r\to\infty))-W(\phi(r=0))|.
\ee
For solutions obeying the first-order equations
\be\label{fosingle}
\frac{d\phi}{dr} =\pm\frac{W_\phi}{r},
\ee
we get to the minimum energy case, with $E_f=E_B$. The total energy is $E=E_B + E_c$, where $E_c = 2\pi A_0(r=0)$ is the energy due to the electric charge. One can show that the above equation is compatible with the equation of motion \eqref{eom1}. We further notice that equations of the above form, with the $1/r$ factor, were considered before in Refs.~\cite{prl,global}, and they engender scale invariance. Moreover, the above first-order equations have two signs; they are related by the change $r\to1/r$, so we only deal with the positive sign from now on. We further notice that equations similar to the above first order equations \eqref{fosingle} appeared before in \cite{internal,research}, for instance, in the study of vortices with internal structure.

The first order equation allows that we write the energy density in Eq.~\eqref{rhofpolar} in the form
\be\label{rhow}
\rho_f = \left(\frac{d\phi}{dr}\right)^2 = \frac{W_\phi^2}{r^2}
\ee
Since $\textbf{E}=-\nabla A_0$, we can use Eqs.~\eqref{esingle}, \eqref{psingle} and \eqref{fosingle} with positive sign to obtain
\be\label{a0wsingle}
A_0 (r) = -W(\phi(r)),
\ee
such that the energy of the charge, $E_c$, is written as
\be\label{ecwsingle}
E_c = 2\pi |W(\phi(0))|,
\ee
and it can be calculated straightforwardly, by just knowing the value of $\phi(r)$ at $r=0$.

Before illustrating the general results, let us notice that if we consider $\varepsilon(\phi)$ such that $\varepsilon(0)=1$, the model \eqref{lmodel} becomes the standard model, with the electric field correctly given by \eqref{esingle} with $\varepsilon(0)=1$ in the standard case. However, when one implements the first order framework to get minimum energy configurations, the first order equations \eqref{fosingle} are mandatory, and they impose that the constant field $\phi=\bar\phi$ has to obey $W_\phi(\bar\phi)=0$, meaning that a constant scalar field configuration cannot be chosen at will anymore. A direct consequence of this appears in Eq. \eqref{psingle}: any constant field $\bar\phi$ that obeys the first order equations \eqref{fosingle} induces a divergence in the dielectric function $\varepsilon(\phi)$, and this imposes that the electric field in Eq. \eqref{esingle} has to vanish in this case to regulate the energy of the system.

Another interesting issue arises from the first order equations \eqref{fosingle}; as it was shown in Ref. \cite{prl}, if one changes $r\to e^x$
the equations \eqref{fosingle} become
\be 
\frac{d\phi}{dx}=\pm W_\phi,
\ee
which are first order equations that describe static configurations of a $(1,1)$ dimensional scalar field theory with potential
\be  
V(\phi)=\frac12 W_\phi^2.
\ee
In this sense, if one chooses 
\be\label{wphin}
W(\phi) = \lambda \,a^2\phi-\frac{1}{3}\,\lambda\, \phi^{3},
\ee
where $\lambda$ and $a$ are real parameters, we get the scalar field model
\be  
{\cal L}=\frac12 \partial_\mu\phi\partial^\mu\phi-V(\phi),
\ee
with the potential 
\be  
V(\phi)=\frac12 \lambda^2 (a^2 -\phi^2)^2,
\ee
and the first order equations 
\be\label{phi4pm} 
\frac{d\phi}{dx}=\pm\lambda(a^2-\phi^2),
\ee
with the standard solutions $\phi(x)=\pm a\tanh[\lambda(x-x_0)]$. The above $(1,1)$ dimensional scalar field model is the prototype of the Higgs field; it is driven by a double-well potential that engenders spontaneous symmetry breaking and has important consequences in nonlinear science. In this sense, the scalar field which drives the symmetry breaking may be seen as an order parameter similar to the spontaneous polarization which is studied, for instance, in ferroelectric materials; see, e.g., Refs. \cite{lgd,Chem,ferro}. We notice that the parameters $\lambda$ and $a$ which we added in \eqref{wphin} control the energy barrier in the double-well potential of the scalar field, so they are of direct interest in applications in ferroelectrics \cite{lgd,Chem,ferro}, for instance. For simplicity, however, we will take $\lambda=a=1$ in this work. 

\begin{figure}
\centering
\includegraphics[width=6.0cm]{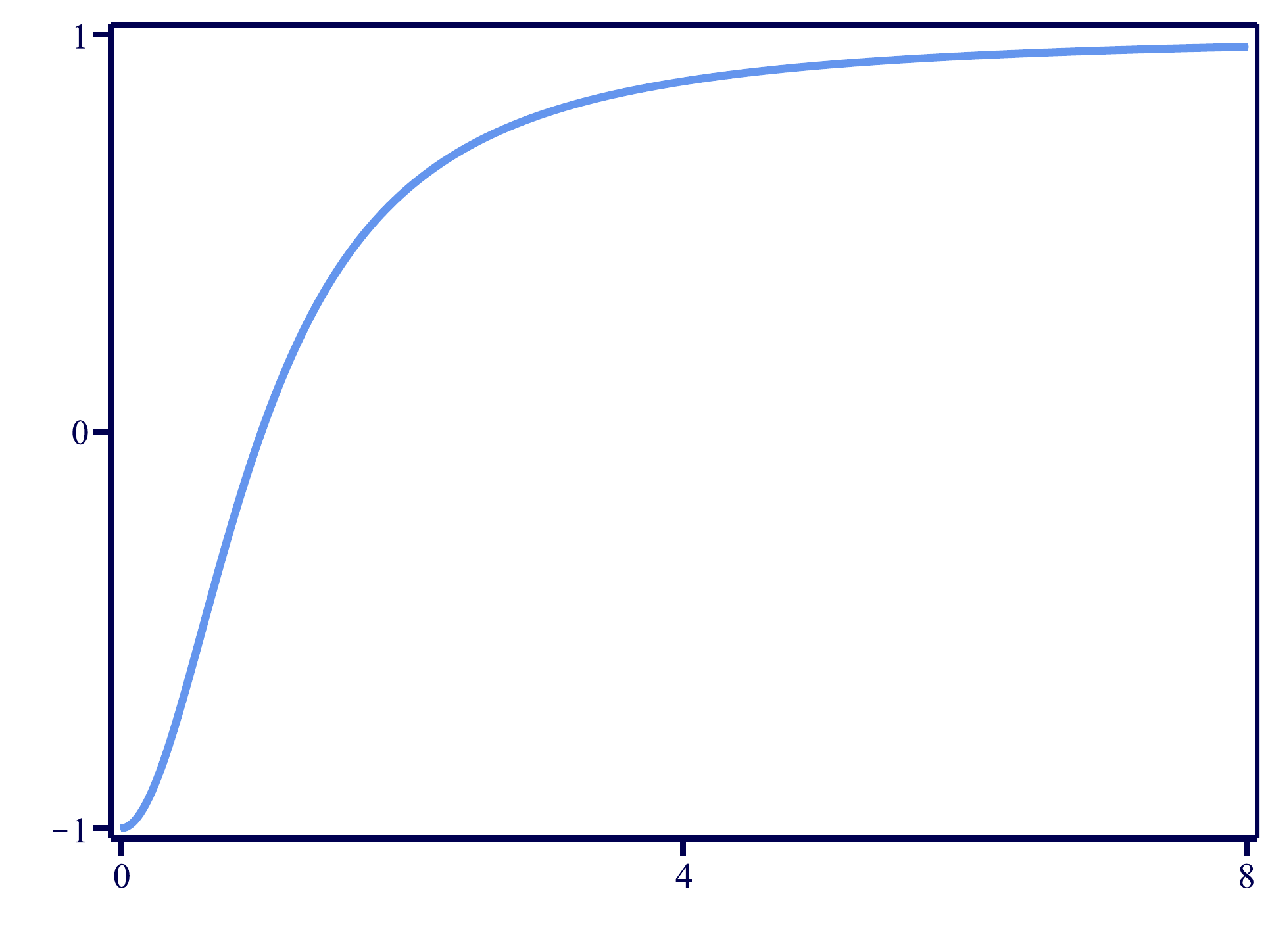}
\includegraphics[width=6.0cm]{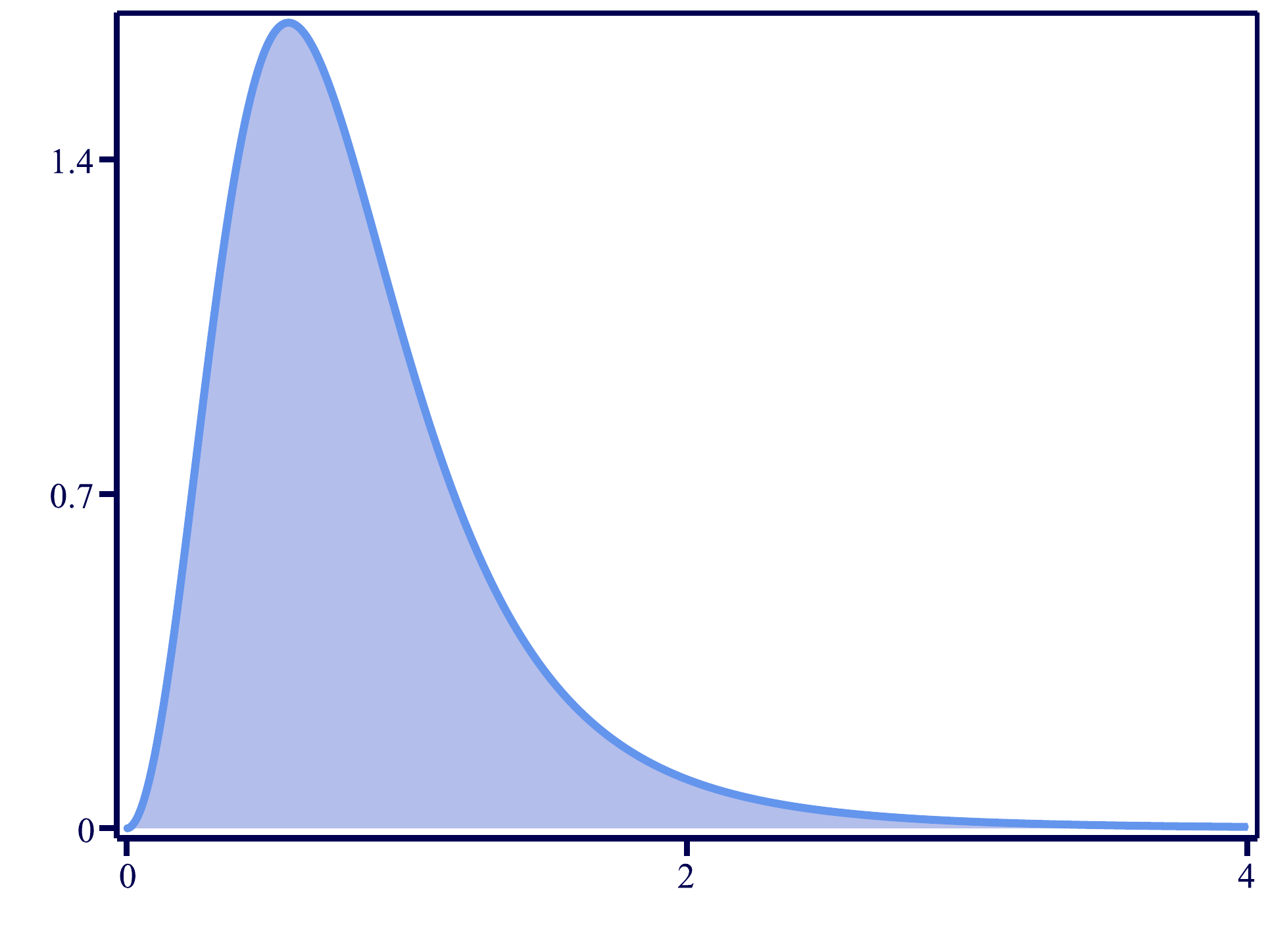}
\caption{The solution $\phi(r)$ that appears in \eqref{sol01} (top) and its energy density $\rho_f(r)$ in \eqref{eneden} (bottom), displayed in terms of the radial coordinate.}
\label{fig1}
\end{figure}

\begin{figure}
\centering
\includegraphics[width=5.0cm]{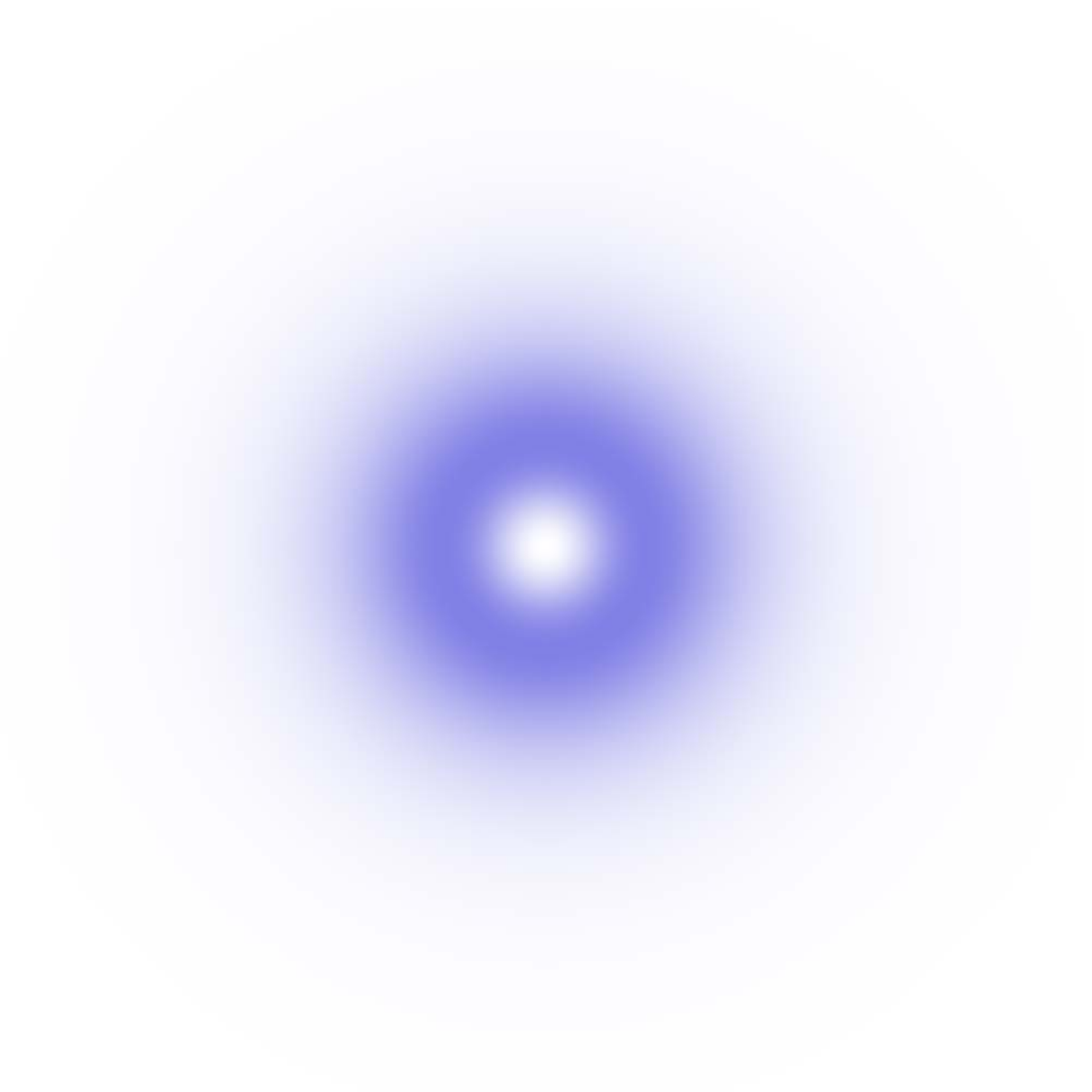}
\caption{The energy density of the solution \eqref{eneden}, displayed with the intensity of the blue color increasing with the increasing values of the energy density.}
\label{fig2}
\end{figure}

Motivated by the above results, let us now illustrate the planar model with $W(\phi)$ as in Eq. \eqref{wphin} (with $\lambda=a=1$). Here the dielectric function in Eq.~\eqref{psingle} is then written as
\be\label{die1}
\varepsilon(\phi) = \left(1-\phi^{2}\right)^{-2},
\ee
and Eq.~\eqref{fosingle} gives
\be\label{fon}
\frac{d\phi}{dr} = \frac{1-\phi^{2}}{r}.
\ee
It is solved by 
\be\label{sol01}
\phi(r) = \frac{r^2-r_0^2}{r^2+r_0^2},
\ee
which has energy density 
\be\label{eneden} 
\rho_f=\frac{16r_0^4r^2}{(r_0^2+r^2)^4}.
\ee
In Fig.~\ref{fig1}, we display the solution of the above first-order equation and its energy density. It engenders scale invariance, so we use the condition $\phi(1)=0$, for simplicity. From Eqs.~\eqref{ebsingle} and \eqref{ecwsingle}, we get that the energy is $E= E_B + E_c$, where $E_B=8\pi/3$ and $E_c= 4\pi/3$, which matches with the numerical integration of the energy density displayed in Fig.~\ref{fig1}. In order to have a better view of the localized solution, in Fig. \ref{fig2} we display the energy density \eqref{eneden} in the $(r,\theta)$ plane. We further notice that it is controlled by the dielectric function $\varepsilon(\phi)$ that couples the gauge field to the scalar field. 

As we have already commented on, the scalar field model described by \eqref{wphin} develops spontaneous symmetry breaking and can be related to the potential based on the phenomenological Landau-Ginzburg-Devonshire or LGD theory of ferroelectrics \cite{lgd,Chem,ferro}. The recent measurements \cite{ferro} of the intrinsic double-well energy landscape in a thin layer of ferroelectric material integrated into a heterostructure with a second dielectric layer, suggest that the negative capacitance \cite{negative} of the dielectric material has its origin in the energy barrier of the double-well potential. The negative capacitance can be used to provide, for instance, voltage amplification for low power nanoscale devices; this can be done, for instance, after trading a standard insulator with a ferroelectric insulator of appropriate thickness \cite{nega}.  


\subsubsection{Second model}

We now focus on the second model. Here, considering the presence of the same electric charge at the origin, the electric field has the same form of Eq.~\eqref{esingle} with $\varepsilon(\phi)\to \varepsilon(\phi,\chi)$, so the equations of motion that describe the scalar fields are
\bes\label{eom2}
\bal
r\frac{d}{dr}\left(r\frac{d\chi}{dr}\right) -\frac{r^2f_\chi}{2}\left(\frac{d\chi}{dr}\right)^2 &= \frac{d}{d\chi}\left(\frac{1}{2\varepsilon}\right),\\
r\frac{d}{dr}\left(rf\frac{d\phi}{dr}\right) &= \frac{d}{d\phi}\left(\frac{1}{2\varepsilon}\right).
\eal
\ees
The energy density for the solutions of the above equations can be written as $\rho = \rho_f + \rho_c$, where
\bes
\bal\label{rhofpolar2}
	\rho_f &=\frac12\left(\frac{d\chi}{dr}\right)^2 +\frac12f(\chi)\left(\frac{d\phi}{dr}\right)^2 + \frac{1}{2r^2\varepsilon(\phi,\chi)},\\
	\rho_c	 &= A_0\frac{\delta(r)}{r}.
\eal
\ees
As before, here we also introduce an auxiliary function $W=W(\phi,\chi)$ to write $\rho_f$ as
\be
\begin{aligned}
	\rho_f &= \frac12\left( \frac{d\chi}{dr} \mp \frac{W_\chi}{r}\right)^2 +\frac12f(\chi)\left( \frac{d\phi}{dr} \mp \frac{W_\phi}{rf(\chi)}\right)^2  \\
		   &+ \frac{1}{2r^2\varepsilon(\phi,\chi)} -\frac{W^2_\chi}{2r^2}-\frac{W^2_\phi}{2r^2f(\chi)} \pm \frac1r\frac{dW}{dr}.
\end{aligned}
\ee
We choose the dielectric function in the form
\be\label{p2}
\varepsilon(\phi,\chi) = \left(\frac{W^2_\phi}{f(\chi)}+W^2_\chi\right)^{-1},
\ee
to make the energy bounded, i.e., $E_f\geq E_B$, where
\be\label{ebsingle2}
E_B = 2\pi |\Delta W|,
\ee
such that $\Delta W = W(\phi(\infty),\chi(\infty))-W(\phi(0),\chi(0))$. One can show that the bound is saturated to $E_f= E_B$ if the following first-order equations are satisfied
\bes\label{fo2}
\bal\label{fo2chi}
\frac{d\chi}{dr} &=\pm \frac{W_\chi}{r},\\ \label{fo2phi}
\frac{d\phi}{dr} &=\pm \frac{W_\phi}{rf(\chi)}.
\eal
\ees
It is not hard to show that solutions of the above first order equations also solve the equations of motion \eqref{eom2}. In this situation, the total energy is $E=E_B+E_c$, where $E_c = 2\pi A_0(r=0)$. The upper and lower signs are related by the change $r\to1/r$, so we only consider the positive sign from now on. By using the above equations, one can write the energy density \eqref{rhofpolar2} as the sum of two contributions, in the form $\rho_f=\rho_1 + \rho_2$, where
\bes\label{rhow2}
\bal\label{rho1}
\rho_1 &= f(\chi)\left(\frac{d\phi}{dr}\right)^2 \\ \label{rho2}
\rho_2 &=\left(\frac{d\chi}{dr}\right)^2.
\eal
\ees
The first-order equations \eqref{fo2} with positive sign may be combined with Eqs.~\eqref{esingle} under the change $\varepsilon(\phi)\to \varepsilon(\phi,\chi)$ and \eqref{p2} to give
\be
A_0(r) = - W(\phi(r),\chi(r)).
\ee
The energy associated to the electric charge is
\be\label{ecw2}
E_c = | W(\phi(0),\chi(0))|.
\ee
We remark that the energy contributions presented in \eqref{ebsingle2} and \eqref{ecw2} do not depend on $f(\chi)$. Also, the function $W=W(\phi,\chi)$ depends on both $\phi$ and $\chi$, so the first order equations \eqref{fo2} are coupled and must be solved simultaneously, in general. However, an interesting case arises for $W(\phi,\chi)= g(\phi) + h(\chi)$. In this specific situation, the first order equation \eqref{fo2chi} can be solved independently, and this simplifies the calculation importantly. This allows that we add another integration constant in the problem, leading to a more general situation. For simplicity, however, we will not consider this possibility in the present work. As in the previous model, here we also have to be careful with the choice of constant fields, $\bar\chi$ and $\bar\phi$, since in the first order framework they have to obey first order differential equations. This issue is similar to the one discussed before in the paragraph below Eq. \eqref{ecwsingle}, so we do not comment on it anymore. 

\begin{figure}[t!]
\centering
\includegraphics[width=6.0cm]{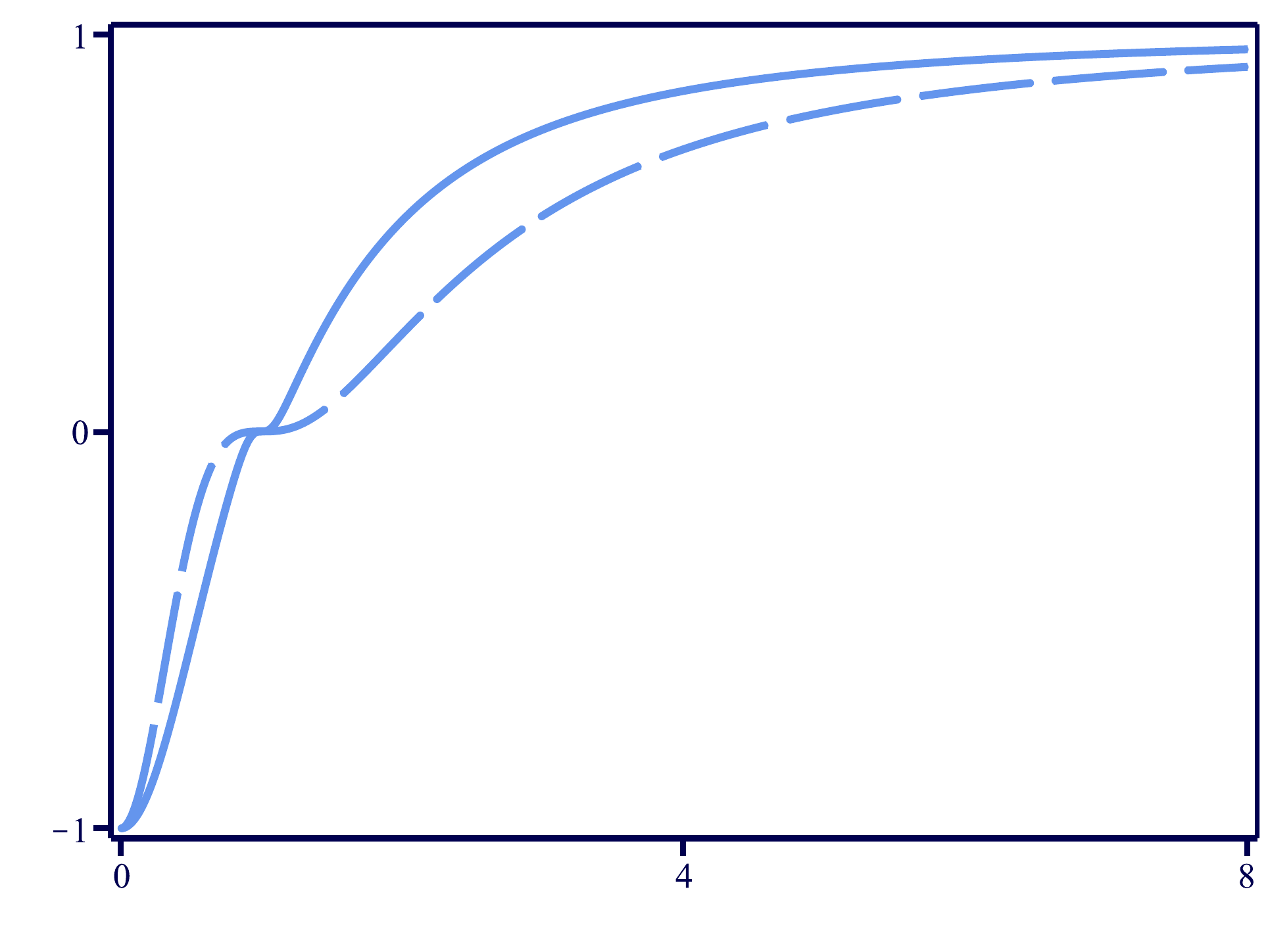}
\includegraphics[width=6.0cm]{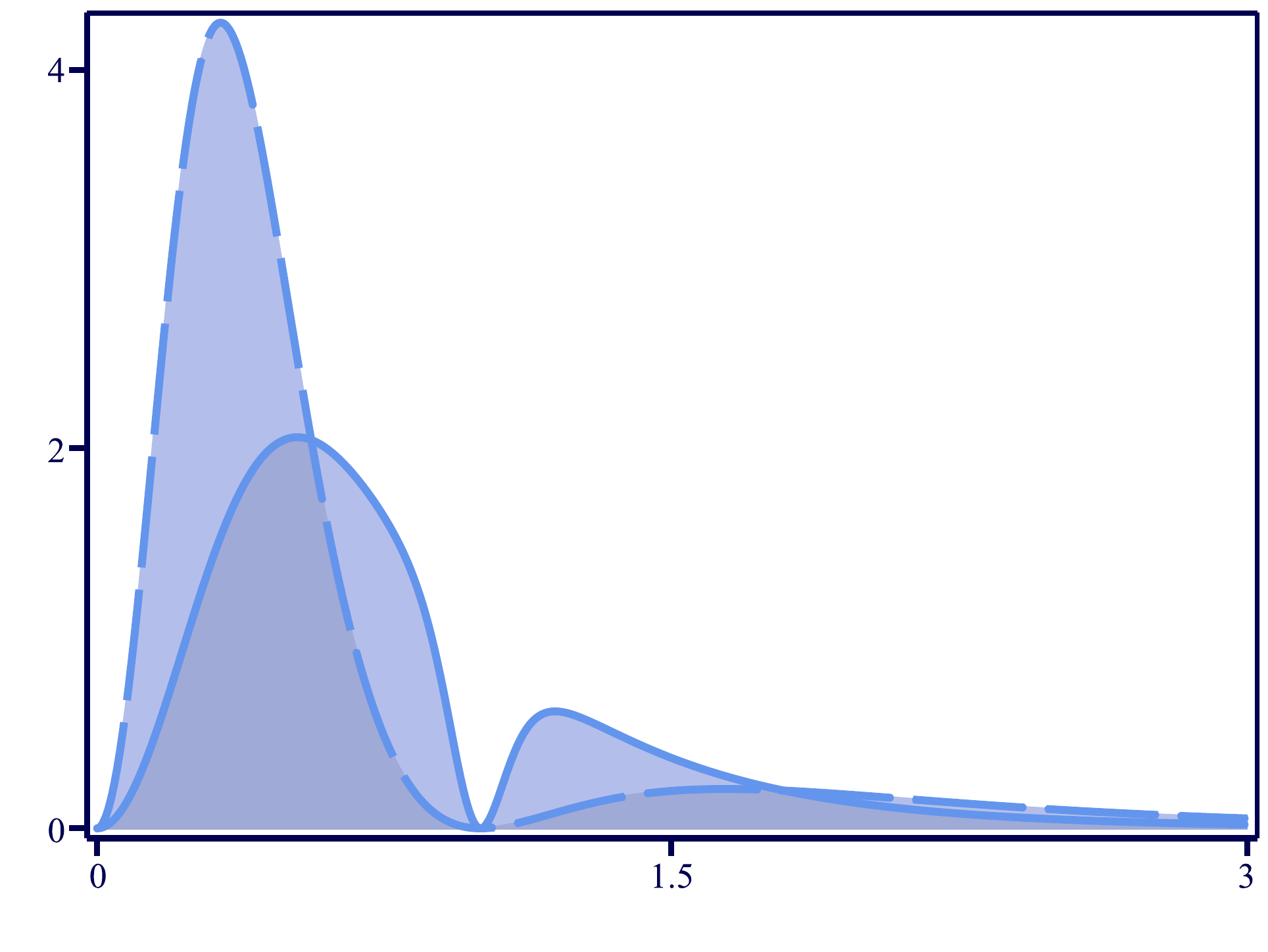}
\caption{The solution $\phi(r)$ in Eq.~\eqref{dk} (top) and its energy density $\rho_1(r)$ in Eq. \eqref{rhodk} (bottom), depicted for $\alpha=2$ (dashed line) and $10$ (solid line).}
\label{fig3}
\end{figure}

\begin{figure}[t!]
\centering
\includegraphics[width=5.0cm]{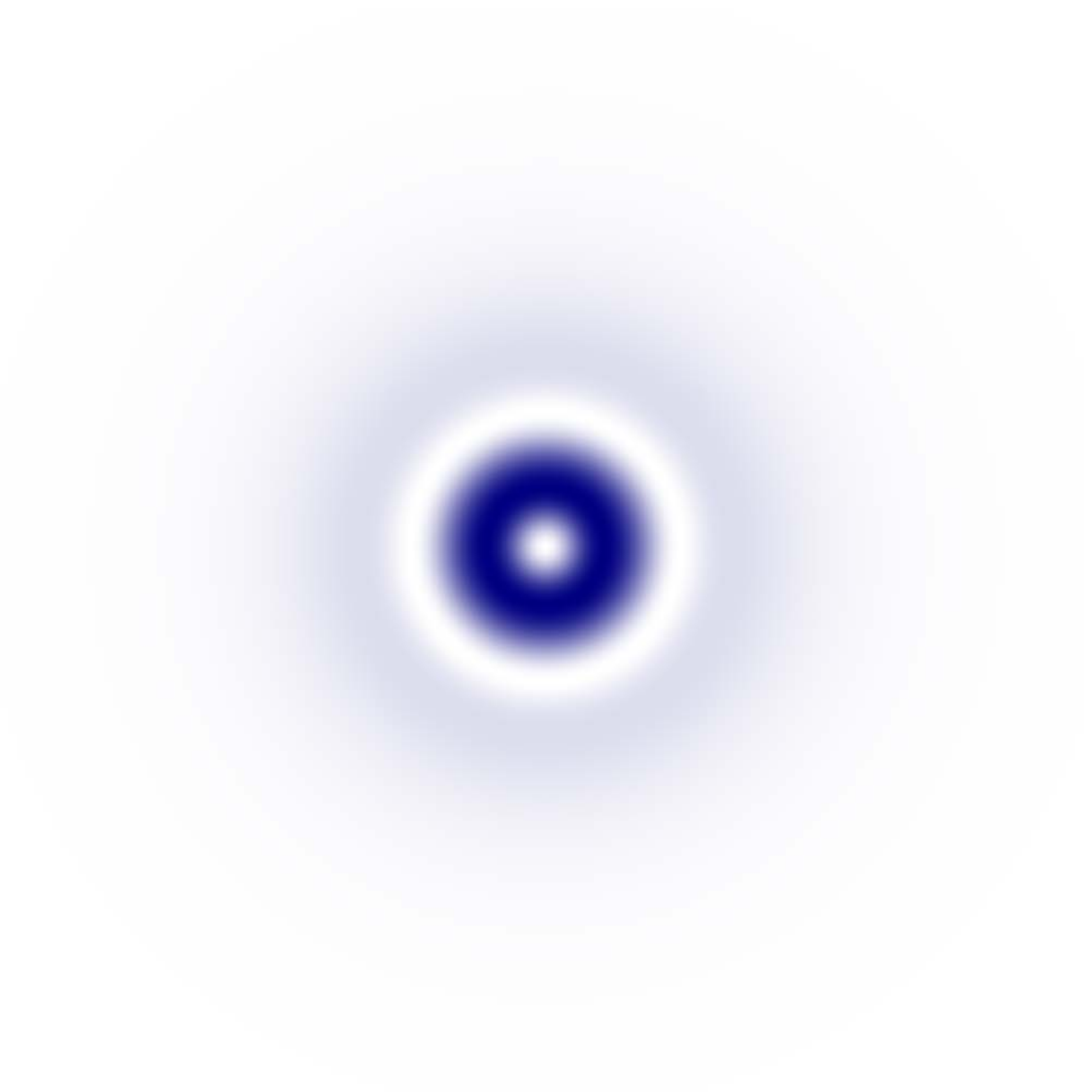}
\includegraphics[width=5.0cm]{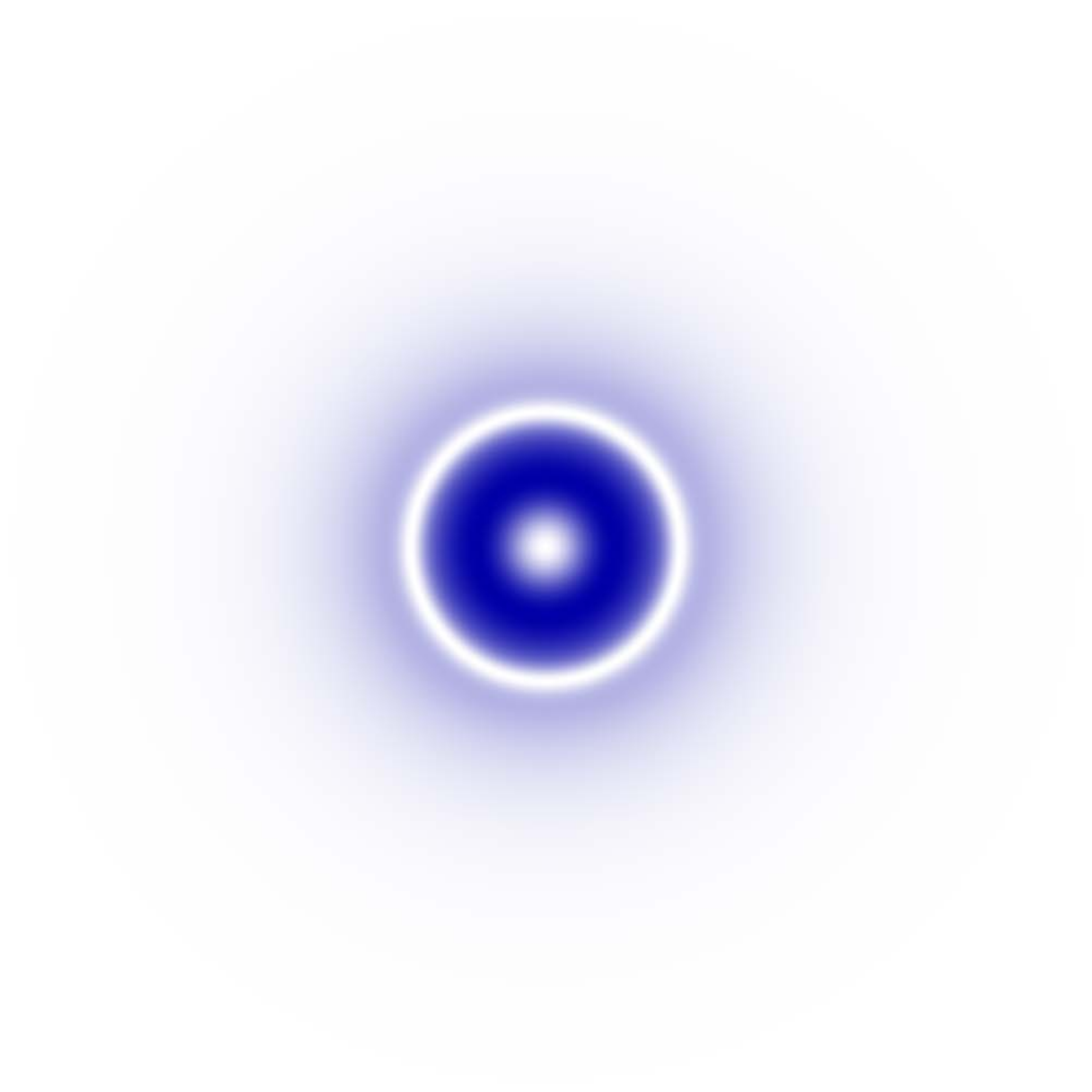}
\caption{The energy density in Eq.~\eqref{rhodk} in the plane for $\alpha=2$ (top) and $10$ (bottom). The intensity of the blue color increases with the increasing of the of the energy densities.}
\label{fig4}
\end{figure}

To illustrate the new possibility, we take
\be\label{h1}
h(\chi)= \alpha\chi - \frac13\alpha\chi^3,
\ee
where $\alpha$ is a positive real parameter. From Eqs.~\eqref{fo2chi} and \eqref{rho2}, we get the solution, $\chi(r)$, and the energy density $\rho_2$, in the form
\be\label{source}
\chi(r) = \frac{r^{2\alpha}-1}{r^{2\alpha}+1} \quad\text{and}\quad\rho_2(r) = \frac{16\alpha^2\,r^{4\alpha-2}}{(r^{2\alpha}+1)^4}.
\ee
Their profiles for $\alpha=1$ can be seen in Fig.~\ref{fig1}, where they appear as the dotted lines. Next, we use the above solution to feed the function $f(\chi)$, which we firstly consider as $f(\chi)=1/\chi^{2}$. This is perhaps the simplest choice, which is not negative, is unity for $\bar\chi=\pm1$, as required by the choice of $h(\chi)$ in Eq. \eqref{h1}, and engenders the appropriate profile. To find the behavior of $\phi(r)$, we choose $g(\phi)$ in the form
\be\label{g1}
g(\phi) = \phi-\frac13\phi^3.
\ee
The dielectric function becomes
\be\label{diefun}
\varepsilon(\phi,\chi)=(\chi^2(1-\phi^2)^2+\alpha^2(1-\chi^2)^2)^{-1},
\ee
and the first order equation \eqref{fo2phi} now changes to
\be
\frac{d\phi}{dr} =\frac{\left(r^{2\alpha}-1\right)^2}{\left(r^{2\alpha}+1\right)^2 r} \left(1-\phi^2\right).
\ee
It supports the analytical solution
\be\label{dk}
\phi(r) = \tanh\left(\ln r - \frac{r^{2\alpha}-1}{\alpha\left(r^{2\alpha}+1\right)} \right).
\ee
The energy density \eqref{rho1} takes the form
\be\label{rhodk}
\rho_1(r) = \frac{\left(r^{2\alpha}-1\right)^2}{\left(r^{2\alpha}+1\right)^2 r^2}\sech^4\left(\ln r - \frac{r^{2\alpha}-1}{\alpha\left(r^{2\alpha}+1\right)} \right).
\ee
The energy of this model is $E = E_B + E_c$, where $E_B = 8\pi(1+\alpha)/3$ and $E_c = 4\pi(1+\alpha)/3$, matching with Eqs.~\eqref{ebsingle2} and \eqref{ecw2}. The solution \eqref{dk} and the above energy density are displayed in Fig.~\ref{fig3}. Notice that the solution presents a plateau around $r=1$ that gets wider as $\alpha$ decreases. This introduces a hole in the energy density at this point, which becomes more evident as $\alpha$ decreases. In order to highlight this feature, we plot the above energy density in the plane in Fig.~\ref{fig4}. The structure presents a hole at the center and a ring around it, which is more visible as $\alpha$ increases. This behavior is controlled by the dielectric function $\varepsilon(\phi)$.


\begin{figure}[t]
\centering
\includegraphics[width=6.0cm]{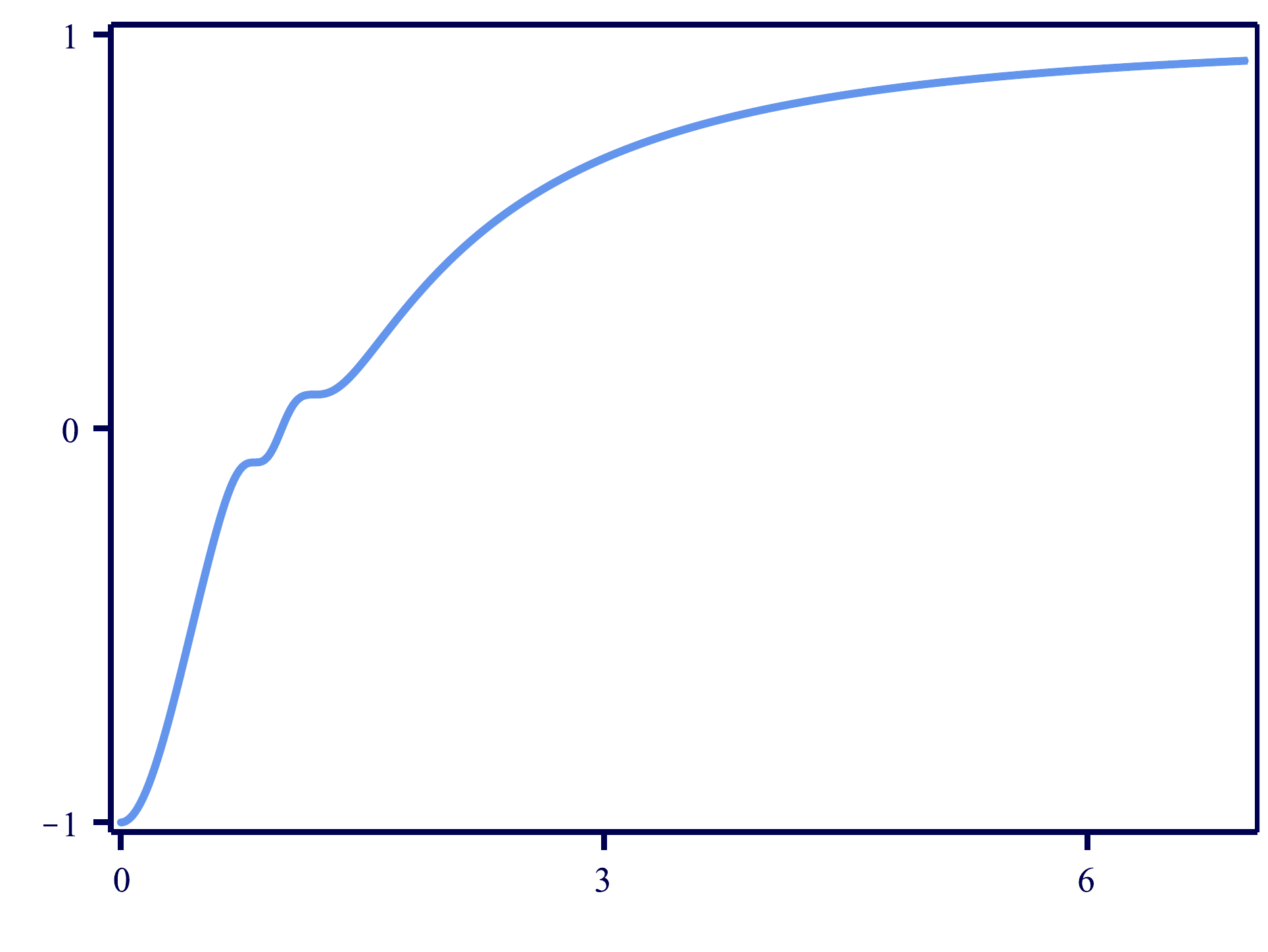}
\includegraphics[width=6.0cm]{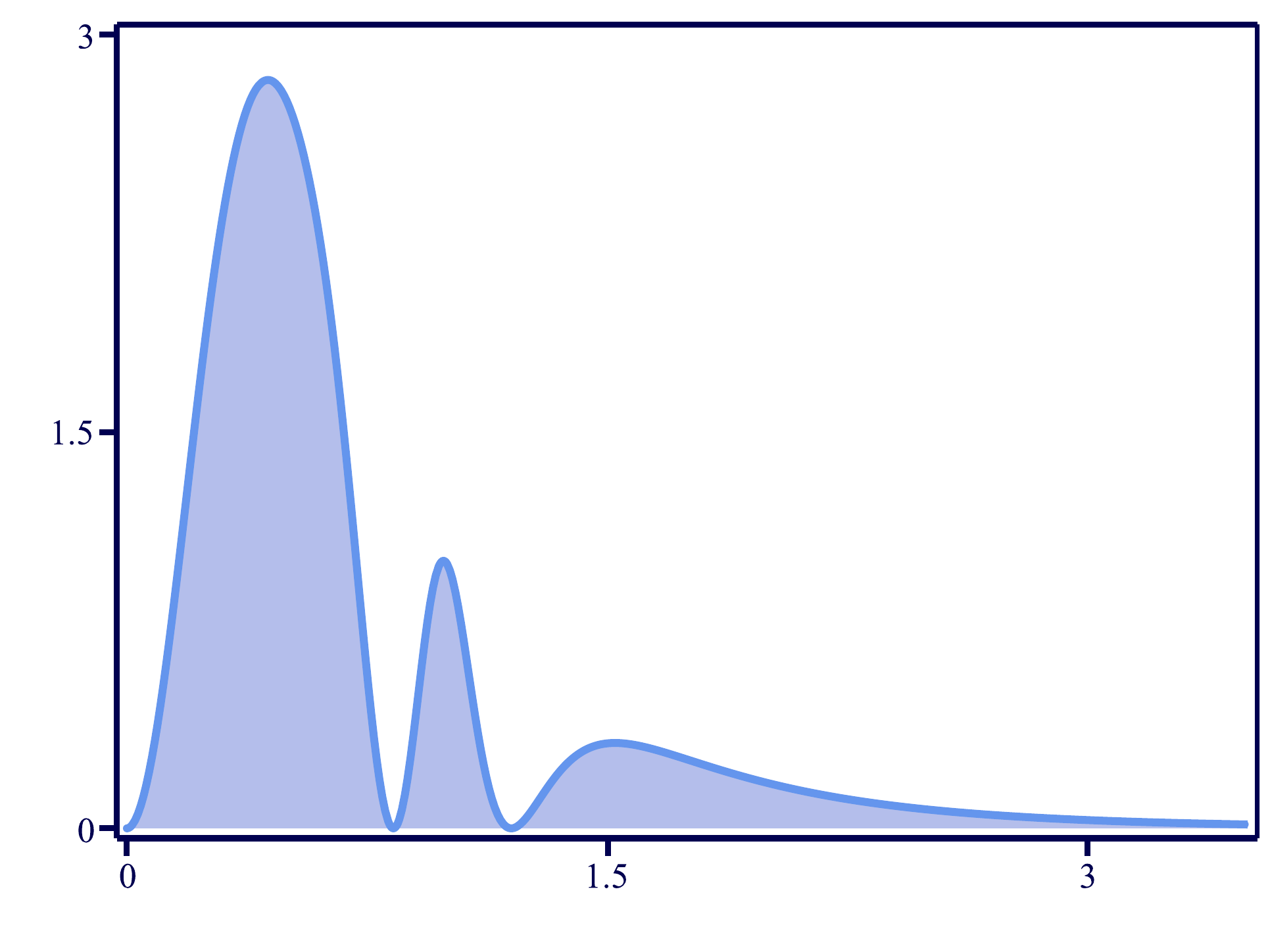}
\includegraphics[width=6.0cm]{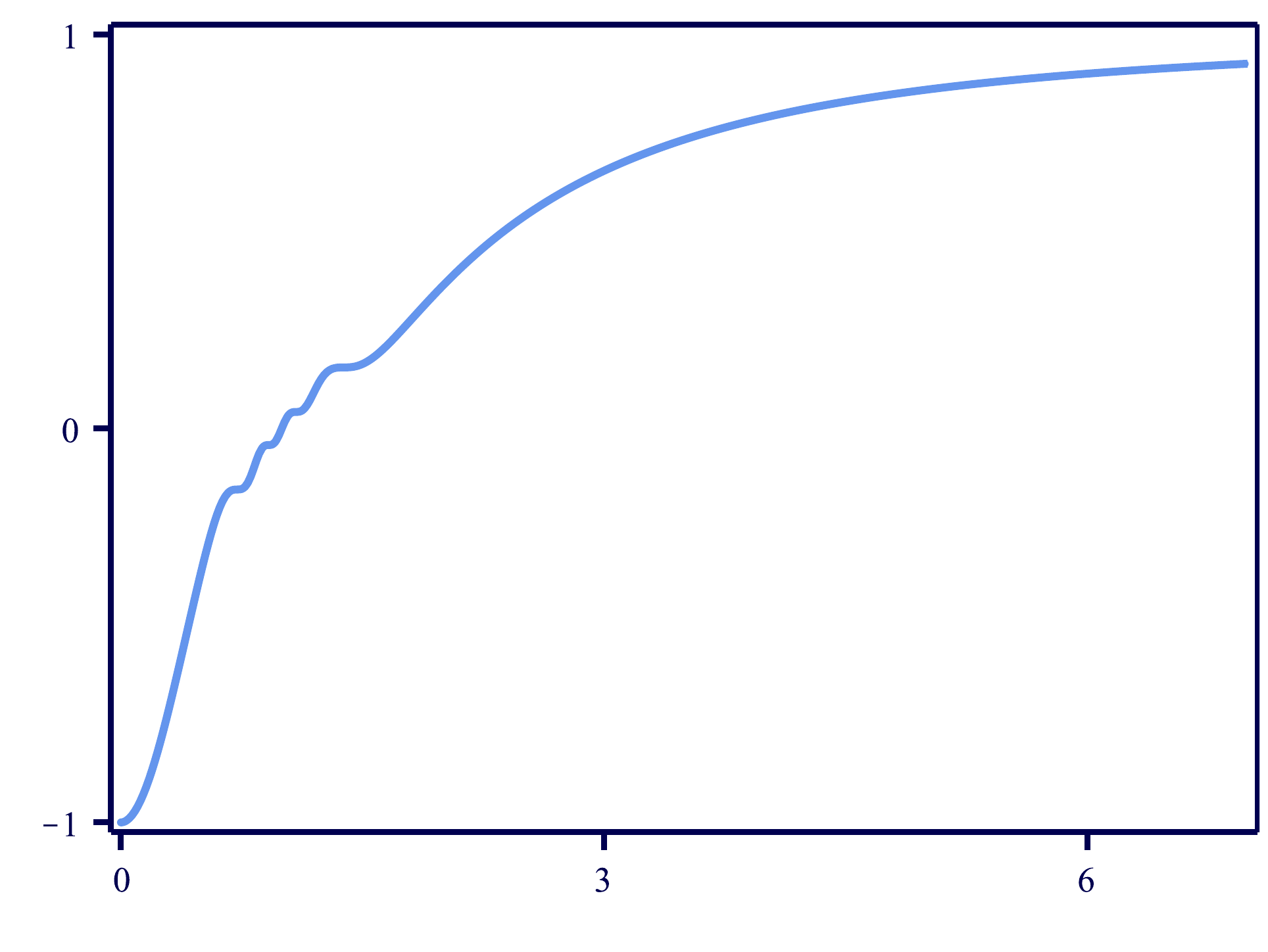}
\includegraphics[width=6.0cm]{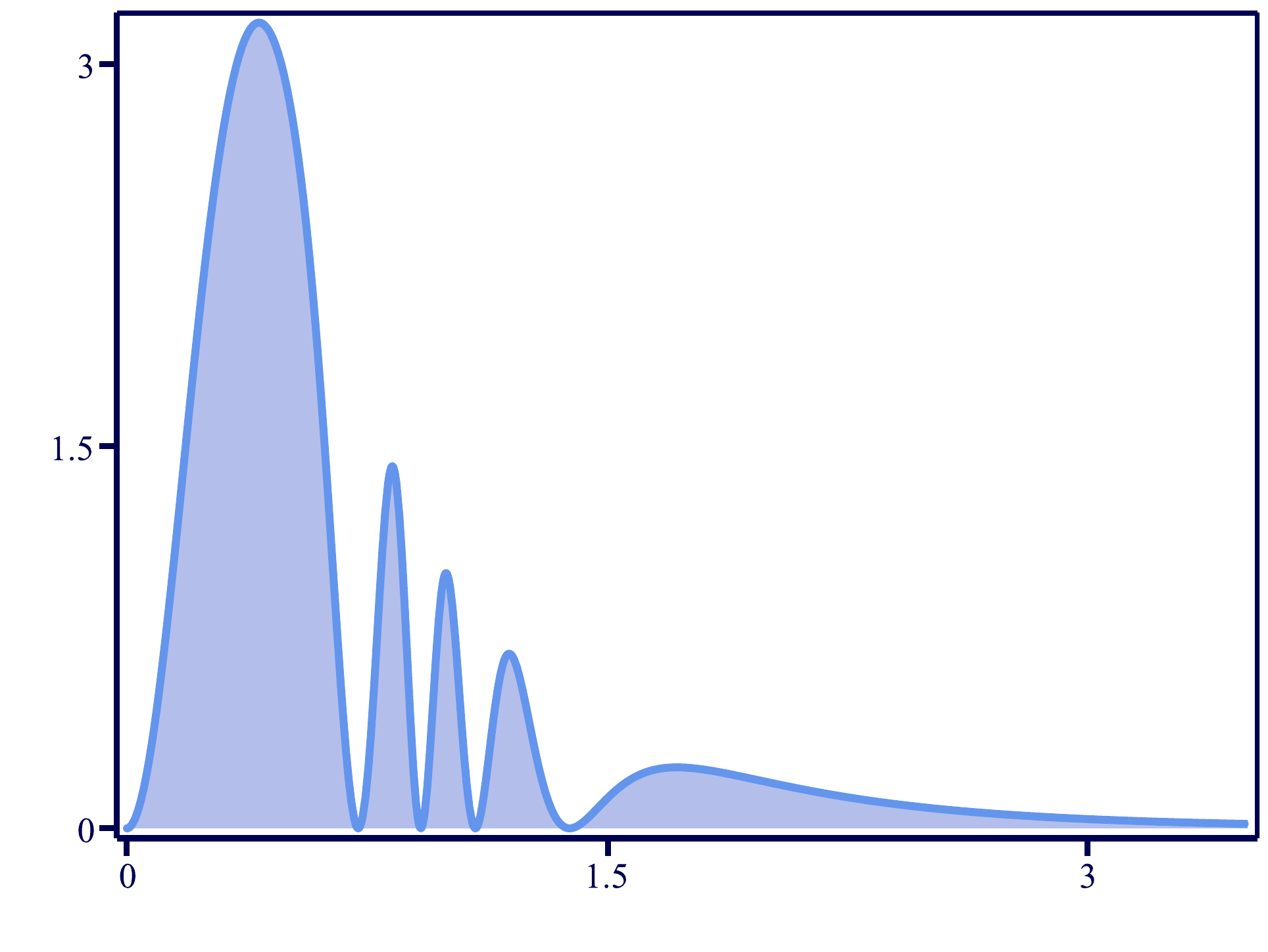}
\caption{The solution in Eq.~\eqref{mk} and its energy density \eqref{rhomk} for $\alpha=3$ and for $n=1$ (top and middle top) and $2$ (middle bottom and bottom), respectively.}
\label{fig5}
\end{figure}

\begin{figure}
\centering
\includegraphics[width=5.0cm]{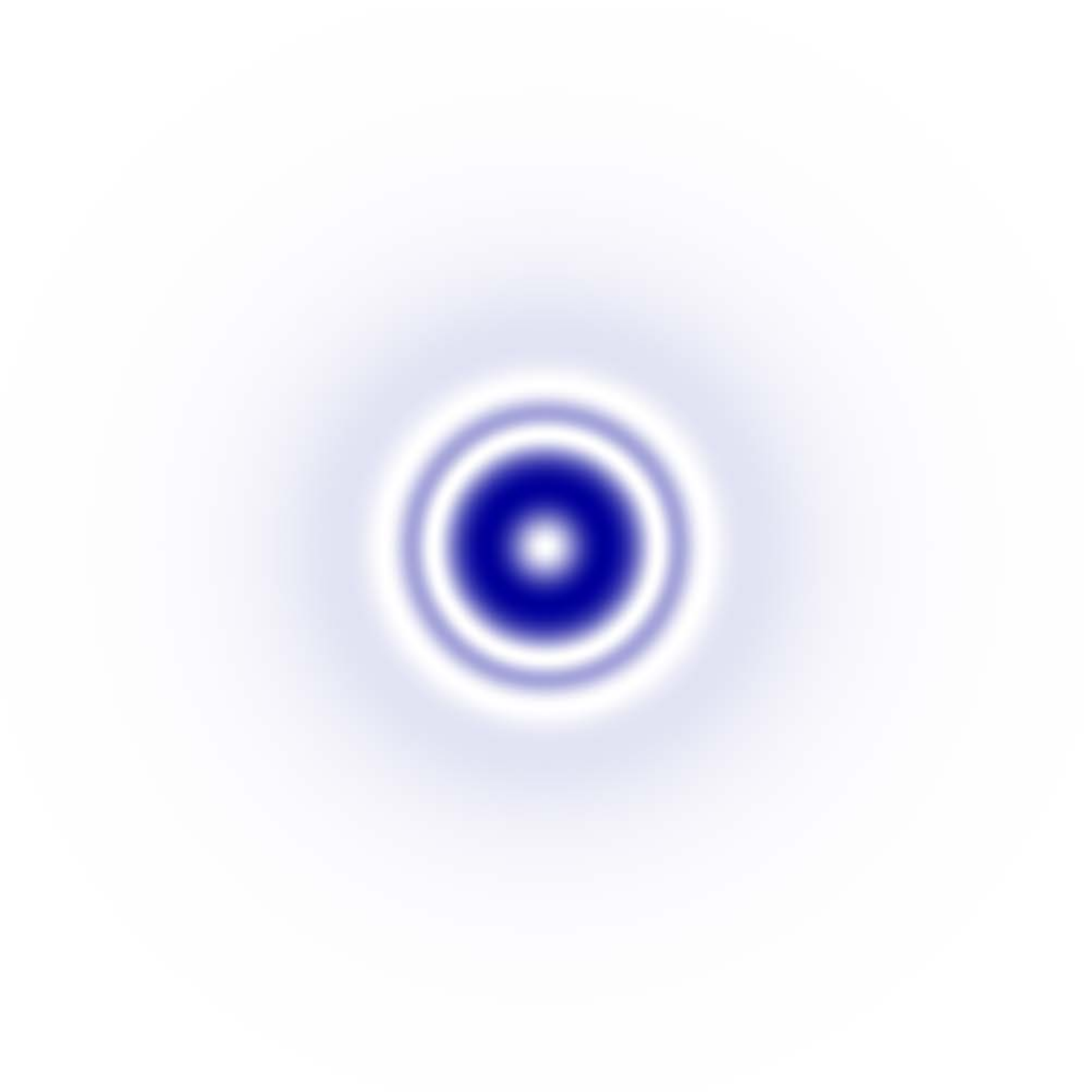}
\includegraphics[width=5.0cm]{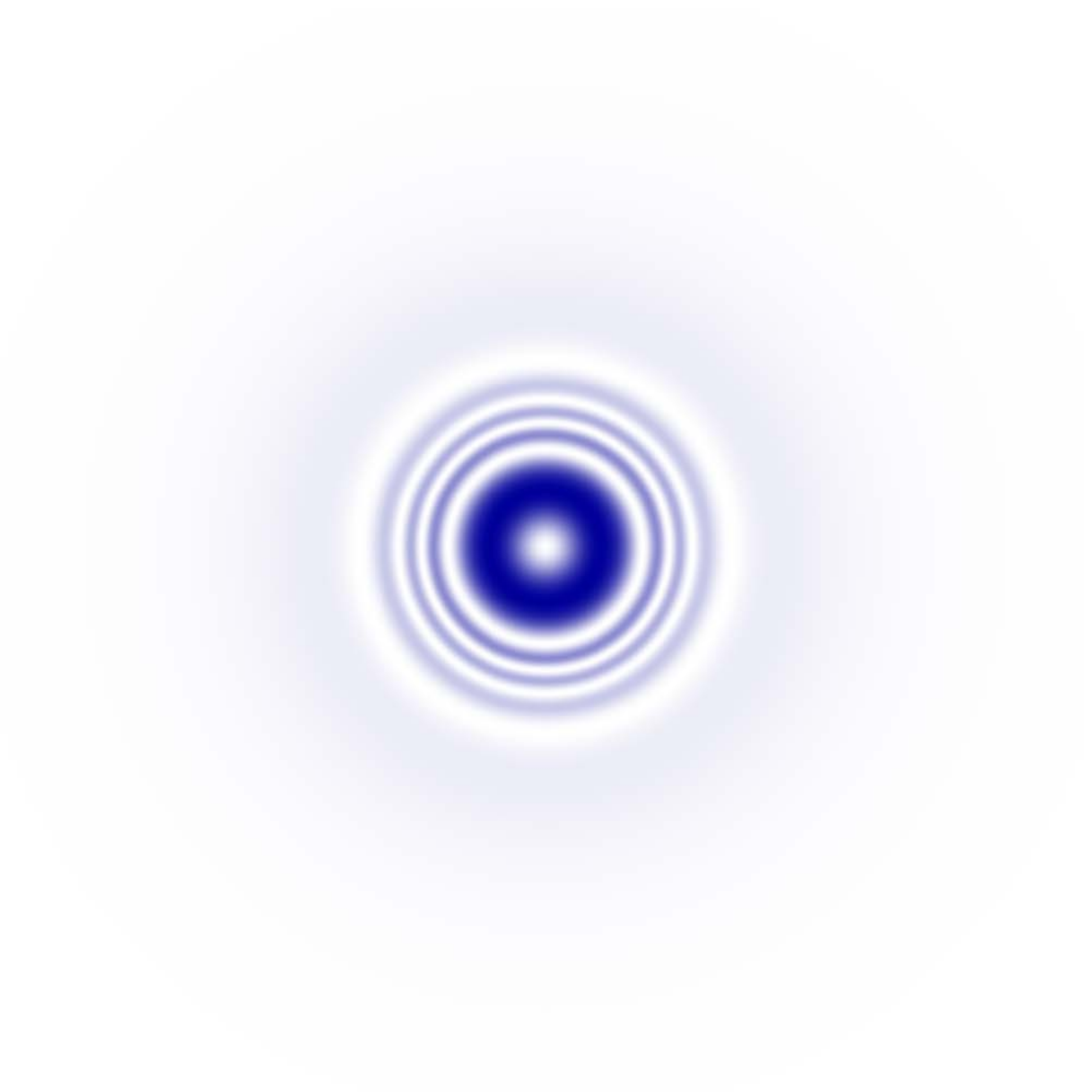}
\caption{The energy density in Eq.~\eqref{rhomk} in the plane for $\alpha=3$ and $n=1$ (top) and $2$ (bottom). The intensity of the blue color increases with the increasing of the energy densities.}
\label{fig6}
\end{figure}

The model described by Eq.~\eqref{g1} with $f(\chi) = \chi^{-2}$ supports rings around the central hole in the energy density. We keep the same $g(\phi)$ in Eq.~\eqref{g1} with the $\chi$ field as in Eq.~\eqref{source}, but now we  consider $f(\chi)=\sec^2(n\pi\chi)$, where $n$ is a natural number. This is another choice, which is also not negative, unity for $\bar\chi=\pm1$, as required by the choice of $h(\chi)$ in Eq. \eqref{h1} above, and engenders the interesting wavelike profile which will bring internal modification in the electric structure. We also remark that, since $W(\phi,\chi)$ does not change, the energy remains the same, i.e., $E = E_B + E_c$, where $E_B = 8\pi(1+\alpha)/3$ and $E_c = 4\pi(1+\alpha)/3$. In this case, we have to change $\chi^2\to \cos^2(n\pi\chi)$ in the factor that multiplies $(1-\phi^2)^2$ in the dielectric function shown in Eq. \eqref{diefun}. Moreover, the first order equation \eqref{fo2phi} takes the form
\be
\frac{d\phi}{dr} =\frac1r\cos^2\left(n\pi\frac{r^{2\alpha}-1}{r^{2\alpha}+1}\right) \left(1-\phi^2\right).
\ee
It supports the solution
\bes\label{mk}
\bal
\phi(r) &= \tanh\vartheta(r),\\
\vartheta(r)&= \frac12\ln{r} + \frac{1}{4\alpha} \left(\text{Ci}\left(\frac{4n\pi r^{2\alpha}}{r^{2\alpha}+1}\right) - \text{Ci}\left(\frac{4n\pi}{r^{2\alpha}+1}\right)\right),
\eal
\ees
where $\text{Ci}(z)$ denotes the cosine integral function. The energy density in Eq.~\eqref{rho1} takes the form
\be\label{rhomk}
\rho_1(r) = \frac{1}{r^2}\cos^2\left(n\pi\frac{r^{2\alpha}-1}{r^{2\alpha}+1}\right)\,\sech^4\vartheta(r).
\ee
Since we have already seen how the parameter $\alpha$ modifies the configurations, we fix $\alpha=3$ and plot the above solution and energy density for $n=1$ and $2$ in Fig.~\ref{fig5}. We see that, as $n$ increases, the solution exhibits more and more plateaux which appear in the energy density as valleys, points in which $\rho=0$. Including the central valley, one gets $2n+1$ valley in the solutions. To illustrate this feature, we plot the above energy density in the plane for the very same values of $\alpha$ and $n$ in Fig.~\ref{fig6}. The localized configuration develops an interesting internal multilayered ringlike structure.

\subsection{Three spatial dimensions}
\label{sec:space}

In the case of three spatial dimensions, we have two distinct possibilities to work. The first one is the case of cylindrical symmetry, with the spatial position described by the vector $(r,\theta, z)$. This can be achieved when one supposes that the electric charge is described by a uniformly charged wire in the $z$ axis. In this case the $z$ dimension in unimportant, and the planar $(r,\theta)$ dimensions describe the planar system studied above. This means that the planar results that we described in Sec. \ref{sec:plane} are also valid in the three dimensional space with axial symmetry. 

The other possibility is to consider spherical symmetry. In this case, we take the current density in the specific form $j^\mu = (\delta(r)/r^2,0,0,0)$ and study the two models described above, in this scenario in the presence of rotational symmetry in three spatial dimensions. 

\subsubsection{First model}

Let us now focus on the presence of electrically charged localized structures in the model \eqref{lmodel} in $(3,1)$ dimensions. Here we first notice that Eqs.~\eqref{eomgen}-\eqref{rho} are also valid in this case, and more, there is no magnetic field. However, in the spatial case the electric field changes to
\be
\textbf{E} = \frac{1}{\varepsilon(\phi)}\frac{\hat{r}}{r^2}.
\ee
In the presence of spherical symmetry, the equation of motion \eqref{eomphi} for the scalar field becomes
\be
\frac{1}{r^2}\frac{d}{d r}\left(r^2\frac{d\phi}{d r}\right) + \frac12 \varepsilon_\phi|\textbf{E}|^2=0.
\ee
Combining this with the above electric field, we obtain
\be\label{eom13d}
r^2\frac{d}{dr}\left(r^2\frac{d\phi}{dr}\right) = \frac{d}{d\phi}\left(\frac{1}{2\varepsilon}\right).
\ee
We use the Eq. \eqref{rho} in order to write the energy density as $\rho = \rho_f + \rho_c$, where
\bes
\bal\label{rhofpolarX}
	\rho_f &=\frac12\left(\frac{d\phi}{dr}\right)^2 + \frac{1}{2r^4\varepsilon(\phi)},\\
	\rho_c	 &= A_0\frac{\delta(r)}{r^2}.
\eal
\ees
As before, the indices $f$ and $c$ stand for field and charge, respectively. Following similar steps of the previous Sec. \ref{sec:plane}, one can write the dielectric function as in Eq.~\eqref{psingle} and show that $E_f = 4\pi |W(\phi(r\to\infty))-W(\phi(r=0))|$ is the minimum energy if the solutions obey the first order equation
\be\label{fo3}
\frac{d\phi}{dr} = \pm\frac{W_\phi}{r^2}.
\ee
We notice that first order equations similar to the above ones appeared before in \cite{mo1,mo2}, in the study of magnetic monopoles with internal structure in three spatial dimensions. 

\begin{figure}
\centering
\includegraphics[width=6.0cm]{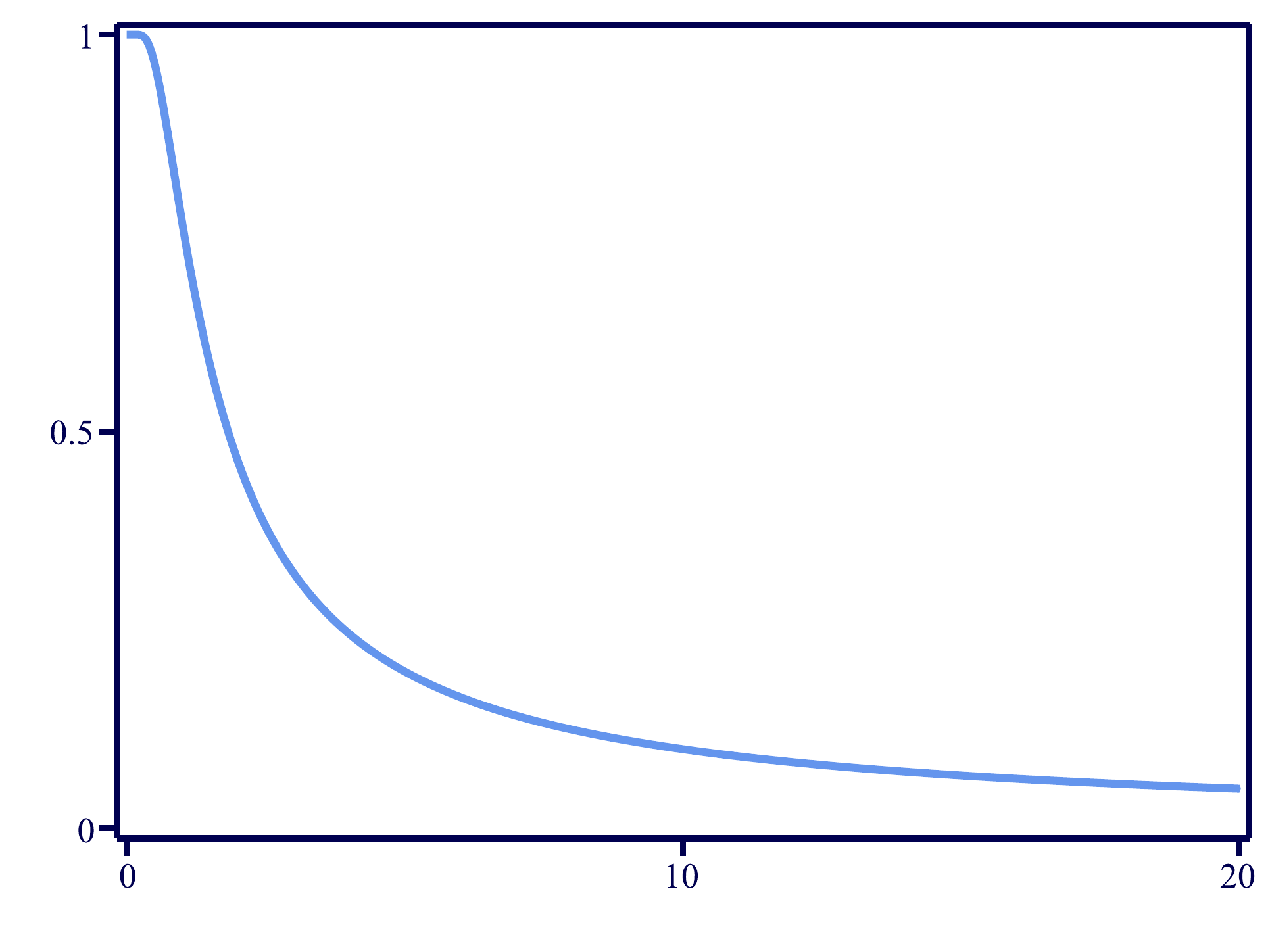}
\includegraphics[width=6.0cm]{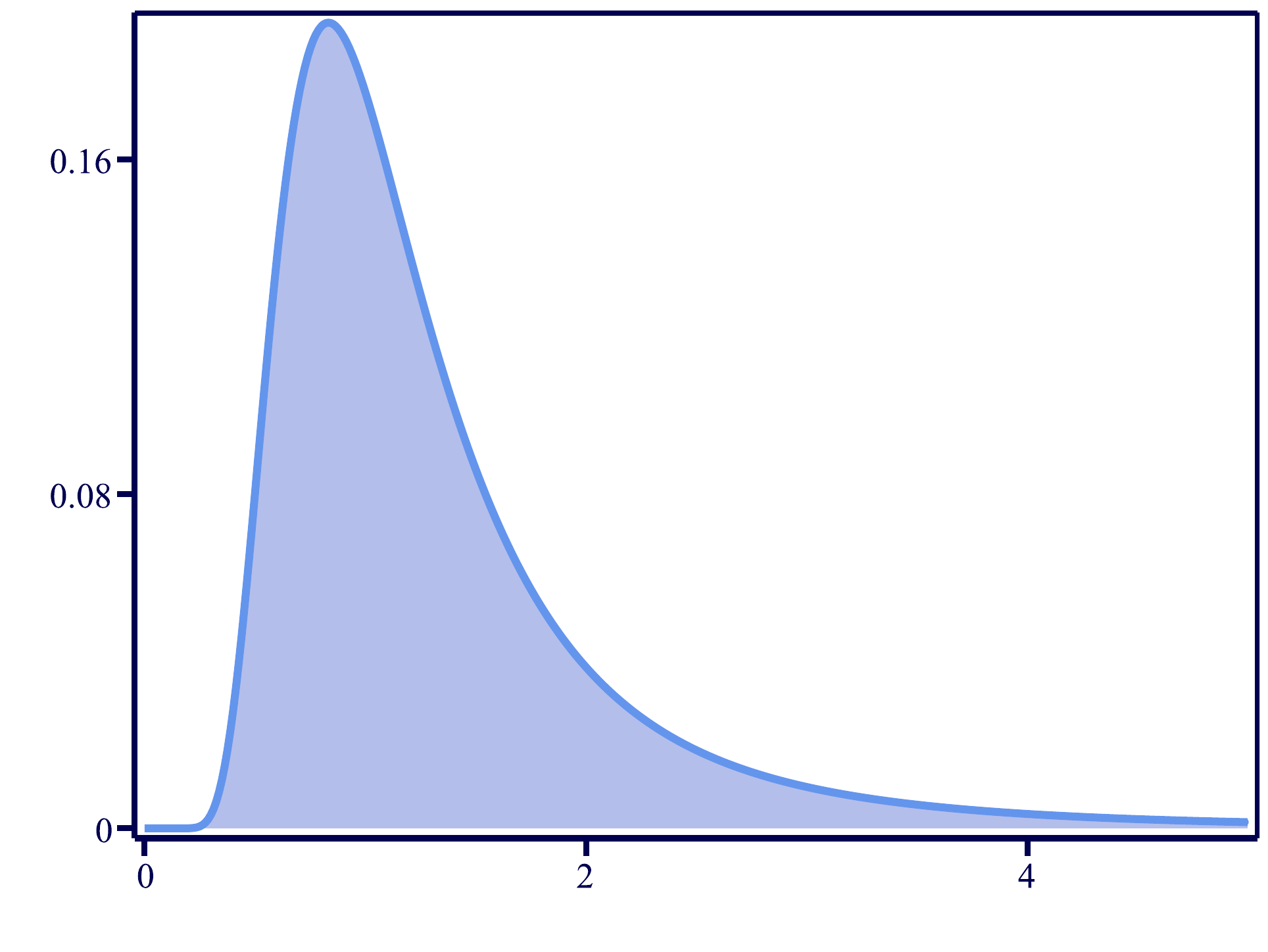}
\caption{The top and bottom panels show the solution and energy density in Eq.~\eqref{phi43d}, respectively.}
\label{fig7}
\end{figure}

Regarding the function $A_0$, it has the very same expression displayed in Eq.~\eqref{a0wsingle}, so $E_c = 4\pi |W(\phi(r=0))|$. Considering the function in Eq. \eqref{wphin}, one can show that the above equation with negative sign,
which we choose for convenience, supports the solution and energy density
\be\label{phi43d}
\phi(r) = \tanh\left(\frac1r\right)\quad\text{and}\quad \rho_f(r) =\frac{1}{r^4}\sech^4\left(\frac1r\right).
\ee
We plot them in Fig.~\ref{fig7}. The planar section of the energy density passing through the center of the structure is similar to the configuration displayed in Fig.~\ref{fig2}, so we do not display it here. One can show that the total energy is $E = E_f + E_c$, where $E_f = E_c = 8\pi/3$.

\subsubsection{Second model}

\begin{figure}
\centering
\includegraphics[width=6.0cm]{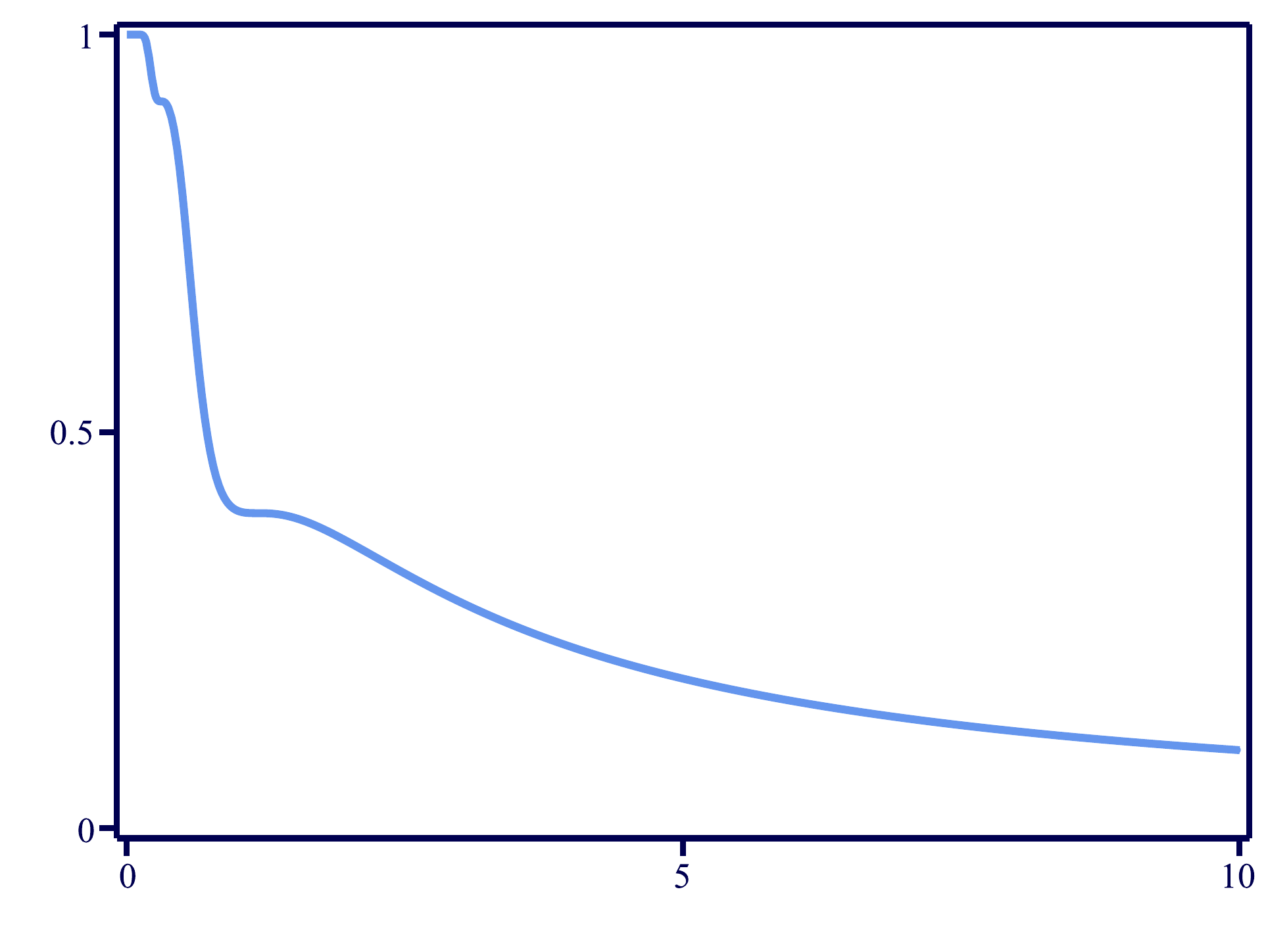}
\includegraphics[width=6.0cm]{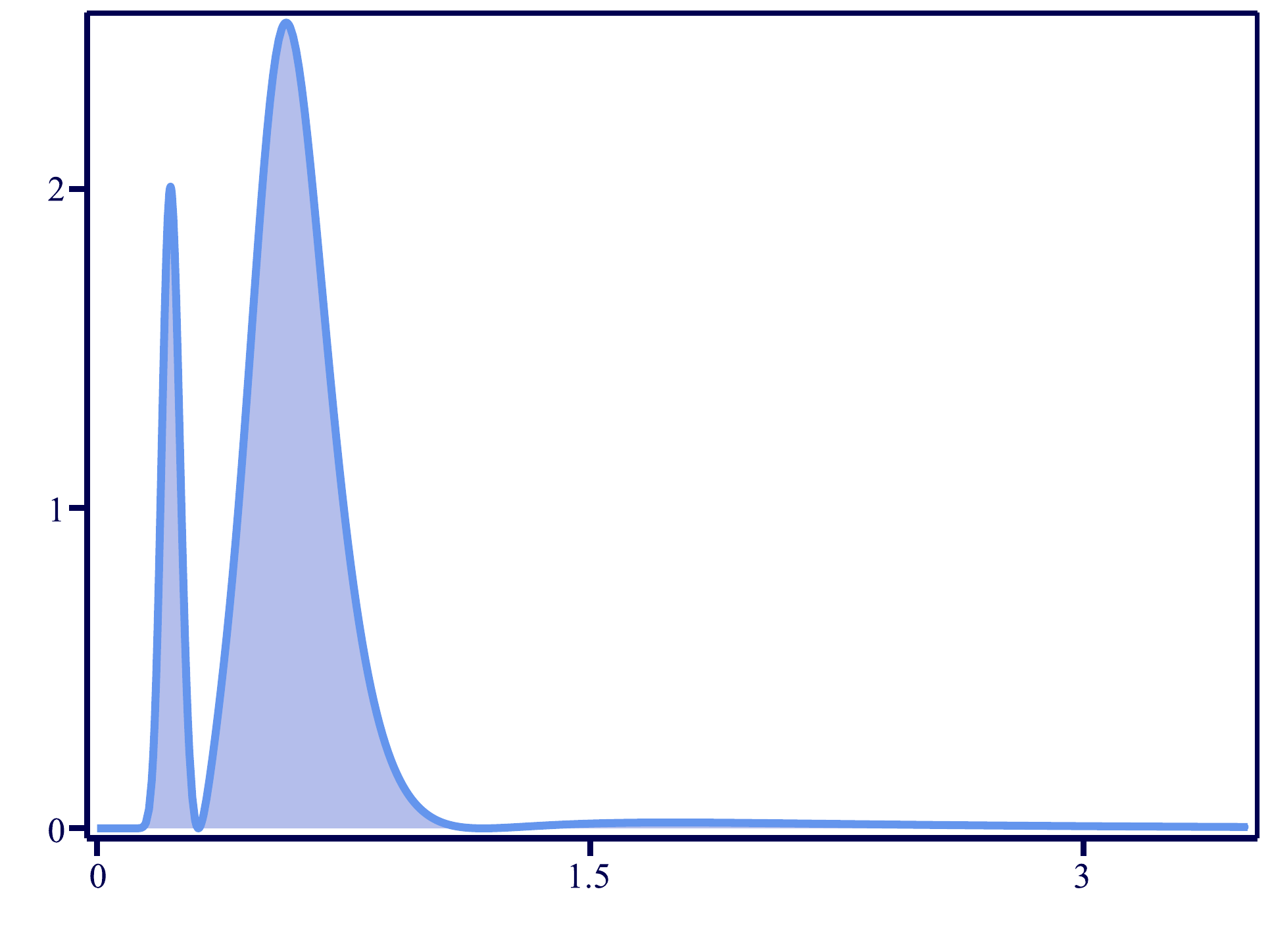}
\caption{The top and bottom panels show the solution \eqref{phi3d} and the the energy density in Eq.~\eqref{rho13d} for $n=2$ and $\alpha=0.3$, respectively.}
\label{fig8}
\end{figure}

\begin{figure}
\centering
\includegraphics[width=5.0cm]{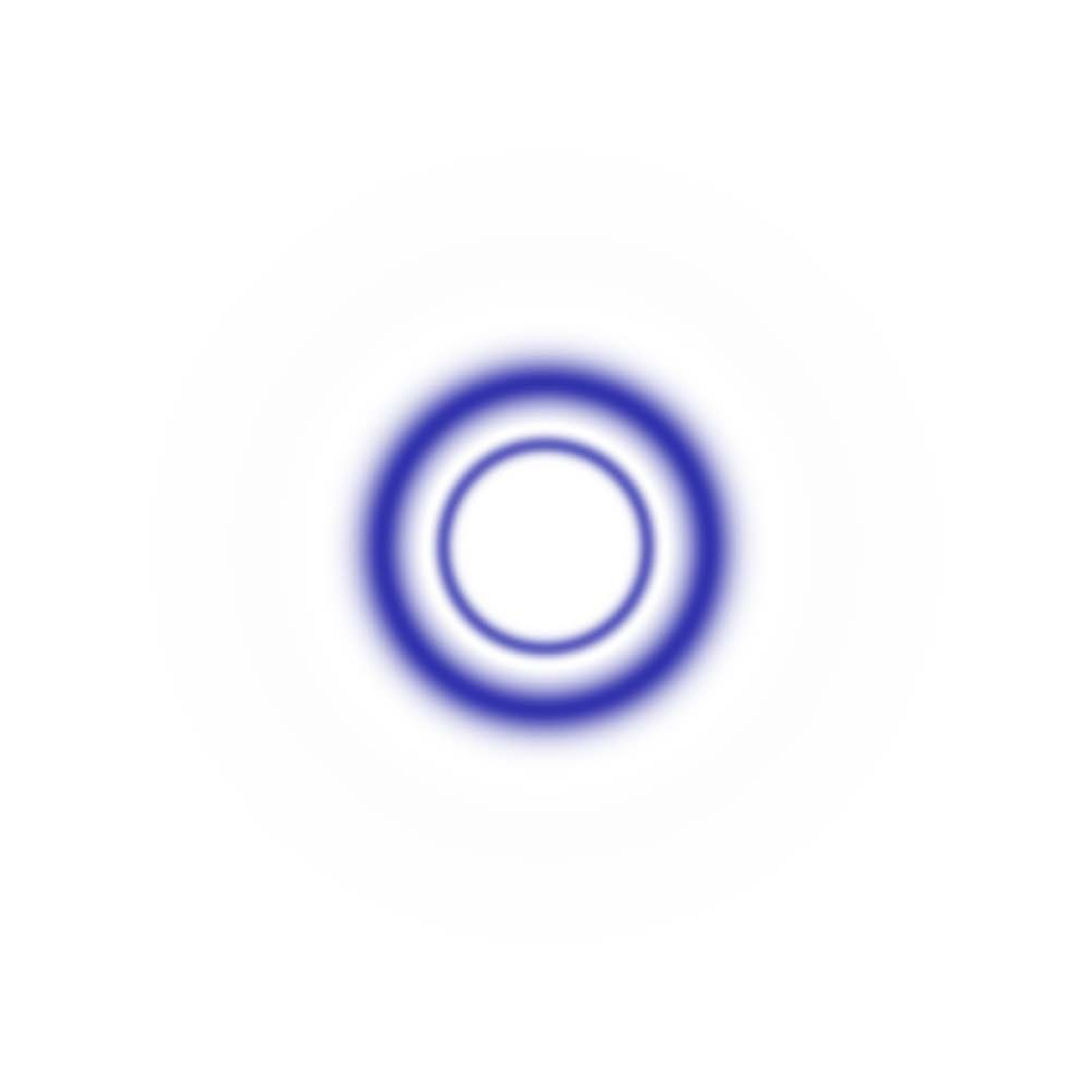}
\caption{The planar section of the energy density passing through the center of the structure for $n=2$ and $\alpha=0.3$.}
\label{fig9}
\end{figure}

We now turn attention to the second model, which is defined in Eq.~\eqref{lmodel2}. We can also develop a first order framework in $(3,1)$ dimensions. In this case, one gets the very same dielectric function in Eq.~\eqref{p2} and, also, the first order equations \eqref{fo2} with the change $r\to r^2$ in the right hand side of the equations, that is,
\bes
\bal
\frac{d\chi}{dr} &=\pm \frac{W_\chi}{r^2},\\ \label{fo2phi}
\frac{d\phi}{dr} &=\pm \frac{W_\phi}{r^2f(\chi)}.
\eal
\ees
Considering the $\chi$ field described by the same $h(\chi)$ in Eq.~\eqref{h1}, we obtain the following solution and associated energy density
\be
\chi(r) = \tanh\left(\frac{\alpha}{r}\right)\quad\text{and}\quad \rho_2(r) = \frac{\alpha^2}{r^4}\sech^4\left(\frac{\alpha}{r}\right).
\ee
They can be seen in Fig.~\ref{fig7} for $\alpha=1$. By using the above solution, one can use the $g(\phi)$ in Eq.~\eqref{g1} with $f(\phi)=\sec^2(n\pi\chi)$ to obtain
\bes\label{phi3d}
\bal
\phi(r) &= \tanh\vartheta(r),\\
\vartheta(r) &= \frac{1}{2r} + \frac{1}{4\alpha}\left(\text{Ci}\left(\xi_+(r)\right) - \text{Ci}\left(\xi_-(r)\right)\right),\\
\xi_\pm(r) &= 2\pi n\left(1\pm\tanh\left(\frac{\alpha}{r}\right) \right),
\eal
\ees
such that its corresponding contribution in the energy density is
\be\label{rho13d}
\rho_1(r) = \frac{1}{r^4}\cos^2\left(n\pi\tanh\left(\frac{\alpha}{r}\right)\right)\sech^4\vartheta(r).
\ee
By using the Bogomol'nyi bound, one gets the energy $E = E_B + E_c$, where $E_B = E_c = 8\pi(1+\alpha)/3$. In Fig.~\ref{fig8}, we display the solution \eqref{phi3d} and the above energy density. We see that the most external ring is almost invisible. In general, the energy density engenders $n+1$ rings, whose intensities are controlled by $\alpha$. To highlight the ringlike profile of the energy density, we plot the planar section of the energy density passing through the center of the structure in Fig.~\ref{fig9}. As both Figs. \ref{fig7} and \ref{fig8} show, the shell structure in the spatial case is different from the ringlike structure in the planar case.


\section{Conclusion}
\label{C}
In this paper, we have studied the effects of a static electric charge immersed in a medium with generalized dielectric function. We have investigated models with a single scalar field, and also with two scalar fields in two and three spatial dimensions. We implemented a first order framework, in which the equations of motion are solved by solutions of first order differential equations, which describe field configurations that minimize the energy of the localized structures. In the case of two scalar fields, we found solutions that allow for the presence of configurations with several internal structures. In both situations, the first order equations simplify the problem and we could find exact analytical solutions in two and three spatial dimensions. 

The models investigated in this work are based on an Abelian gauge field with $U(1)$ symmetry, so it would be of current interest to investigate the possibility to extend the present study to other cases, as in the Born-Infeld \cite{BI,BI2} modification of the Maxwell term that leads to nonlinear electromagnetism, and also in the case of non Abelian gauge fields. The case of an SU(2) gauge theory appears of current interest here, since this is close to the case of magnetic monopoles \cite{M1,M2,mo1,mo2}. Another possibility refers to the case of Q-balls in the presence of $U(1)$ gauged two-component model similar to the models recently investigated in Refs.~\cite{Q1,Q2}. The dielectric modification that we included in this work may suggest further research in the subject.

We notice that first order equations similar to the equations \eqref{fosingle} appeared before in \cite{internal,research} in the study of vortices with internal structure in the planar case. Moreover, in three spatial dimensions, equations of the first order type similar to the equations \eqref{fo3} also appeared before in \cite{mo1,mo2} in the investigation of magnetic monopoles. It is interesting to see that these first order equations change as one changes from two to three spatial dimensions, but the change is directly related to the study developed in \cite{prl}, in which we investigated the presence of topological structures constructed by a single real scalar field in arbitrary dimensions, circumventing the scaling theorem due to Derrick and Hobart \cite{H,D}, which shows that there is no topological structure in dimensions greater than one when one deals with standard scalar field theory. We recall here that in Ref. \cite{prl} one changed the potential of the scalar field, adding specific spatial dependence which allowed to construct topological solutions in arbitrary dimensions. In this sense, the investigation in \cite{prl} seems to provide a unifying approach to deal with the presence of localized structures described by real scalar fields in two and three spatial dimensions.

The effect responsible for the presence of internal structure appears from the requirement of finite energy of the localized structure. Indeed, for higher and higher values of the dielectric function, the electric field has to diminish toward zero to keep the energy finite. In this sense, if one thinks of applications of the present study to other areas of nonlinear science, one would require materials with high dielectric constants, which can be found, for instance, in ceramic elements like the ones investigated in Refs. \cite{D1,D2,D3}. There are other possibilities as the ones investigated, for instance, in \cite{G1,G2}; in \cite{G1} the study dealt with dielectric film with a high dielectric constant using chemical vapor deposition-grown graphene interlayer and in \cite{G2} the authors investigated the construction of dielectric gels with a new type of polymer-based dielectric material in order to design gels that achieve ultra-high values for the dielectric constant. In the case of three spatial dimensions with axial symmetry, the results of the planar case may suggest the study of optical fibers with an electrically charged ultra thin wire encapsulated at the core of the fibers and other possible realizations. Moreover, in the case of three spatial dimensions, the localized structure can be seem as core and shell nanoparticles in the form of polymer based nanocomposite dielectrics that create hierarchically structured composites in which each sublayer may contribute a distinct function to yield a multilayered multifunctional material \cite{Science, nanotech} similar to the core and shell magnetic structures described in Ref. \cite{magnetic}, which appears as the magnetic counterpart of the electrically charged structures described in the present work.   
 
\acknowledgements{This work is supported by the Brazilian agencies Conselho Nacional de Desenvolvimento Cient\'ifico e Tecnol\'ogico (CNPq), grants Nos. 404913/2018-0 (DB), 303469/2019-6 (DB) and 306504/2018-9 (RM), and Paraiba State Research Foundation (FAPESQ-PB) grants Nos. 0003/2019 (RM) and 0015/2019 (DB and MAM).}

\end{document}